\documentclass[12pt,preprint]{aastex}

\slugcomment{Last Modified \today}

\shorttitle{Correlations between MIR, FIR, H$\alpha$, and FUV Luminosities for SWIRE galaxies}
\shortauthors{Zhu et al.}

\begin{document}

\title{Correlations between Mid-Infrared, Far-Infrared, H$\alpha$, and FUV Luminosities for $Spitzer$ SWIRE-field Galaxies}
\author{Yi-Nan Zhu\altaffilmark{1,2}, Hong Wu\altaffilmark{1}, Chen Cao\altaffilmark{1,2,3}, Hai-Ning Li\altaffilmark{1,2}}
\altaffiltext{1}{National Astronomical Observatories, Chinese Academy of Sciences, Beijing 100012, China; zyn@bao.ac.cn; hwu@bao.ac.cn; lhn@bao.ac.cn} 
\altaffiltext{2}{Graduate University, Chinese Academy of Sciences, Beijing 100039, China}
\altaffiltext{3}{Institute of Space Science and Physics, Shandong University at Weihai, Weihai, Shandong 264209, China; ccao00@gmail.com}

\begin{abstract}
We present and analyze the correlations between mid-infrared (MIR), far-infrared (FIR), total-infrared (TIR), H$\alpha$, and FUV 
luminosities for star-forming galaxies, composite galaxies and AGNs, based on a large sample of galaxies selected from the $Spitzer$ 
SWIRE fields. The MIR luminosities of star-forming galaxies are well correlated with their H$\alpha$, TIR and FUV luminosities, and 
we re-scaled the MIR-derived SFR formulae according to the above correlations with differences less than 15\%. We confirm the recent 
result by \citet{calzetti07} that the combined observed H$\alpha$ $+$ 24$\mu$m luminosities L(H$\alpha$$_{\rm obs}$$+$ 24$\mu$m) 
possess very tight correlation with the extinction-corrected H$\alpha$ luminosities L(H$\alpha$$\_$corr) for star-forming and 
even for dwarf galaxies, and show that the combined L(H$\alpha$$_{\rm obs}$$+$ 8$\mu$m[dust]) are also tightly correlated with 
L(H$\alpha$$\_$corr) for the above sample galaxies. Among all the L(MIR)-L(FIR) correlations for star-forming galaxies, 
the L(24$\mu$m) vs. L(70$\mu$m) and L(8$\mu$m[dust]) vs. L(160$\mu$m) are the tightest and also nearly linear. The former could 
be related to young massive star formation, while the latter might be relevant to diffuse dust emissions heated by old stellar 
populations. Composite galaxies and AGNs have higher MIR-to-H$\alpha$/MIR-to-FUV luminosity ratios than star-forming galaxies, 
nevertheless their correlations among MIR, FIR and TIR luminosities are completely following those of star-forming galaxies.

\end{abstract}


\section{Introduction}
Since the successful launch of $Infrared~Astronomical~Satellite$ ($IRAS$) in 1983, more advanced space-based infrared telescopes 
had been launched, such as $Infrared~Space~Observatory$ \citep[$ISO$;][]{kessler96} and $Spitzer~Space~Telescope$ \citep{werner04}, 
to help as to study the MIR to FIR emission properties of galaxies. MIR to FIR emissions are crucial for quantifying extragalactic 
star formation activities, since infrared emissions are arising from the dust re-radiation of UV photons emitted by young massive 
stars. FIR luminosity has been proved to be a good star formation rate (SFR) indicator of galaxies by $IRAS$ observations 
\citep[e.g.,][]{hunter86, lehnert96}. However, the low sensitivity and spatial resolution of FIR observations obstructed us from 
applying it to most galaxies at intermediate and high redshifts, even though there are some degree of improvements in the $ISO$ 
and $Spitzer$ era. Optical/UV SFR tracers, such as H$\alpha$-line emission and FUV continuum luminosities, are important because 
they are still the only way to measure SFR in regions of low extinction such as faint, dwarf galaxies and especially extended 
UV disk (XUV-disk) galaxies \citep{thilker07}. But these tracers usually suffer from serious extinction effects for normal and 
especially infrared-bright galaxies which are difficult to be corrected \citep{jonsson04}. MIR emissions of normal galaxies are 
dominated by dust continuum from very small grains (VSGs), together with some broad emission features \citep{gillett73, willner77} 
which were realized to be from polycyclic aromatic hydrocarbons \citep[PAHs;][]{leger84, puget89}. Early analysis of correlations 
between MIR luminosity (derived from $ISO$ observations) and other SFR tracers \citep[e.g.,][]{elbaz02, roussel01, schreiber04, 
flores04} have shown that the MIR luminosity could be used to estimate SFRs of galaxies.

Mid-Infrared observations by $Spitzer$ with higher sensitivity and better angular resolution than $ISO$, provide a new opportunity 
to study young stellar populations and star formation processes in galaxies. The 8$\mu$m band of $Spitzer$ Infrared Array Camera 
\citep[IRAC;][]{fazio04} is designed to cover the 7.7$\mu$m PAH feature; whilst the 24$\mu$m band of Multiband Imaging Photometer 
for $Spitzer$ \citep[MIPS;][]{rieke04} just covers the dust continuum from VSGs for local galaxies, avoiding the contamination of 
most of the MIR emission or absorption lines. The pioneering work by \citet{wu05} (hereafter Wu05) has shown that, for star-forming 
galaxies, both $Spitzer$ IRAC 8$\mu$m(dust) and MIPS 24$\mu$m luminosities are strongly correlated with the 1.4~GHz radio and 
H$\alpha$ luminosities. Then the new formulae for estimating SFRs of galaxies were derived from MIR luminosities based on above 
correlations. Detailed studies of nearby galaxies M~51 \citep{calzetti05} and M~81 \citep{perez06} revealed that the 24$\mu$m 
luminosities of star-forming regions in the above two galaxies were well correlated with the extinction-corrected Pa$\alpha$ luminosities. 
Based on the 8$\mu$m and 24$\mu$m data for HII regions in 33 nearby SINGS \citep[$Spitzer$ Infrared Nearby Galaxies Survey;][]{kennicutt03} 
galaxies, \citet{calzetti07} found that both 8$\mu$m(dust) and 24$\mu$m luminosities are correlated with L(Pa$\alpha$$\_$corr), 
but non-linearly. They and \citet{kennicutt07} also demonstrated that the combination of observed H$\alpha$ (uncorrected for dust 
extinctions) and 24$\mu$m luminosities could be one of the best SFR indicator for HII regions in these galaxies. 

Besides being heated by young massive stars, the energy source of MIR and FIR emissions could also be the active galactic nuclei (AGN) 
in the galactic center, since AGNs are also powerful UV emitters. Weak AGN-hosting galaxies are very popular in the local universe 
\citep{ho97}, thus the possible AGN contribution to MIR and FIR emissions could have significant impact on the estimation of cosmic 
SFR density and the understanding of its evolution \citep[see, e.g,][]{bell05,perez06,elbaz07,daddi07}. Previous works showed that 
powerful AGNs can strengthen MIR VSG emissions, but weaken or even avoid PAH emissions \citep{weedman05,siebenmorgen04,verma05}. 
Weak AGNs are found to have MIR properties different from that of star-forming galaxies \citep{wen07,li07}, and the degree of such 
differences could be related to the AGN activity \citep{wu07}.

The $Spitzer$ Wide-area Infrared Extragalactic Survey \citep[SWIRE;][]{lonsdale03} is the largest extragalactic survey program among 
the six $Spitzer$ cycle-1 Legacy Programs. With a total field of $\sim$49 deg$^{2}$, it is much larger than the $Spitzer$ First Look 
Survey (FLS) field ($\sim$3.7 deg$^{2}$) studied by \citet{wu05}. In this paper, we perform statistical analysis on the correlations 
between MIR, FIR, H$\alpha$, and FUV luminosities for star-forming galaxies, composite galaxies (which were affected by both central 
nuclear and star formation activities), and AGNs selected from three northern SWIRE fields.

The structure of the paper is as follows. We describe the construction of our sample and the estimation of multi-wavelength luminosities 
in $\S$2 and $\S$3. The major results on correlation analysis are presented in $\S$4. Some discussions and a summary of this work 
are given in $\S$5 and $\S$6. Throughout this paper, we adopted a $\Lambda$CDM cosmology with $\Omega_{\rm m}=0.3$, $\Omega_{\rm \Lambda}=0.7$ 
and $H_{\rm 0}=70\,{\rm km \, s^{-1} Mpc^{-1}}$.

\section{Sample and Data Reduction}

\subsection{SDSS Sample}
The optical spectral sample galaxies are selected from the Sloan Digital Sky Survey \citep[SDSS;][]{york00} main galaxy sample 
\citep{strauss02}, with $r$-band Petrosian magnitudes less than 17.77 mag. The SDSS main galaxy sample covers three northern $Spitzer$ 
SWIRE fields: most of the Lockman Hole (center: 14 41 00, +59 25 00) and ELAIS-N2 (center: 16 36 48, +41 01 45), and about 1/3 
of ELAIS-N1 (center: 16 11 00, +55 00 00). The overlay regions used in this paper is about $15$ deg$^{2}$. A total of $821$ galaxies 
were selected with available H$\alpha$ and H$\beta$ emission line fluxes ($> 5$$\sigma$), based on the emission line catalog 
of SDSS DR4 \footnote{http://www.mpa-garching.mpg.de/SDSS/DR4/; see also \citet{tremonti04}.}.

\subsection{MIR Sample}
The IRAC four bands (3.6, 4.5, 5.8 and 8.0~$\mu$m) images were mosaicked from the Basic Calibrated Data (BCD; acquired from the 
$Spitzer$ Sciences Center) after flat-field corrections, dark subtraction, linearity and flux calibrations \citep{fazio04,huang04}, 
with the final pixel scale of 0.6$\arcsec$ \citep{wu05,cao07,cao08}. The MIPS 24$\mu$m images were mosaicked as the same way but 
with the pixel scale of 1.225$\arcsec$ \citep{wen07, cao07, wu07}. Based on the catalogs of the Two Micron All Sky Survey 
\citep[2MASS;][]{cutri03}, we obtained the accuracy of the astrometric calibration around 0.1$\arcsec$ in all the five bands.

We obtained the $auto$ and $aperture$ magnitudes for all the sources in the three SWIRE fields detected by SExtractor \citep{bertin96}. 
The $auto$ magnitudes were used to denote the $total$ luminosities of detected sources. Hereafter, the $auto$ magnitudes were represented 
by the $total$ magnitudes for these five MIR bands. Aperture photometries with 3$\arcsec$ and 6$\arcsec$ diameters were adopted for the 
IRAC four bands and MIPS 24$\mu$m band images, respectively. 

The aperture corrections for the IRAC four band photometries (corrected to an aperture of 24$\arcsec$) are: $-$0.66, $-$0.64, $-$0.80 
and $-$1.00, respectively. We re-calibrated the IRAC $aperture$ magnitudes with the aid of 2MASS photometric catalog, according to 
\citet{lacy05} and \citet{eisenhardt04}. Firstly, we selected blue 2MASS field stars (F7 dwarfs) with $J-K$ color less than 0.3. To 
ensure accurate photometry and avoid saturated sources in IRAC 3.6$\mu$m images, the K-band magnitudes were restricted within 10 to 
14 mags; and then, the measured AB magnitudes in each IRAC band were converted to Vega magnitudes by adding factors of 2.79, 3.26, 
3.73 and 4.40 \footnote{http://spider.ipac.caltech.edu/staff/gillian/cal.html}, respectively; afterwards, we compared the four $K-IRAC$ 
colors with `$Expected~Value$' from \citet{bessel88} and adopted the averaged values of 0.04, -0.04, -0.03 and 0.07 (based on Table~\ref{tab1}) 
as the corrections for each band. The final flux calibration uncertainties of all the four IRAC bands are less than 0.08 mag. 
The 6$\arcsec$ aperture photometry of MIPS 24$\mu$m band was corrected to an aperture of 30$\arcsec$ , with an aperture correction 
value of $-$1.21. Following the procedure used by the MIPS instrument team to derive calibration factors from standard star observations, 
we corrected an additional calibration factor of 1.15. A factor of 0.96 was divided in order for $\nu$F$_{\nu}$ equating constant scale
\footnote{http://swire.ipac.caltech.edu/swire/astronomers/publications/SWIRE2$\_$doc$\_$083105.pdf}. The final MIPS 24~$\mu$m flux has 
a calibration uncertainty less than 10\% \citep{rieke04}.

To construct the MIR galaxy sample, we cross-matched these MIR sources with the above SDSS sample galaxies with a radius of 2$\arcsec$. 
Only $15$ out of the $821$ SDSS galaxy are mismatched. Most of these sources are edge-on galaxies or have bright neighbors. The completeness 
of the MIR sample is $\sim$98\%, since the sensitivity of $Spitzer$ MIR bands is much deeper than that of SDSS spectroscopic observations. 
For the purpose of this work, we need both $total$ and $aperture$ magnitudes in at least four MIR bands: IRAC 3.6, 5.8, 8~$\mu$m, and MIPS 
24$\mu$m. Therefore, another $201$ galaxies were rejected because of their locations at the edge of overlay regions of SDSS and/or $Spitzer$ 
images. The final number of our MIR sample galaxies is $605$.

\subsection{FIR Sample}
The $Spitzer$ MIPS 70 and 160$\mu$m band photometry catalogs of SWIRE northern fields were derived from the SWIRE data release 3 
\footnote{http://swire.ipac.caltech.edu/swire/astronomers/data$\_$access.html}. The 70$\mu$m catalog was cross-matched with the 
160$\mu$m catalog with a radius of 11$\arcsec$, and then cross-matched with the MIR sample with a radius of 4$\arcsec$. Since we 
need to calculate the total infrared (TIR) luminosity, only the sources with detectable fluxes in all three MIPS bands were selected. 
The final FIR sample includes $197$ galaxies. The point-response function (PRF) magnitudes were adopted as total magnitudes in this work, 
and the absolute flux uncertainties are 20\% and 30\% for the two bands, respectively
\footnote{http://swire.ipac.caltech.edu/swire/astronomers/publications/SWIRE2$\_$doc$\_$083105.pdf}. 

\subsection{UV Sample}
The $Galaxy~Evolution~Explorer$ \citep[$GALEX$;][]{martin05} General Data Release 2 (GR2) and General Data Release 3 (GR3) 
\footnote{http://galex.stsci.edu/GR2/} have been released to public in spring 2006 and summer 2007. Two UV bands were used: 
the far-UV (FUV) band covering $\sim$ 135-175 nm and the near-UV (NUV) band covering $\sim$ 175-275 nm. The UV catalogs were 
taken from pipeline-processed GALEX Deep Imaging Survey (DIS) \citep[m$_{AB}$$\sim$25;][]{martin05} observations of three northern 
SWIRE fields. The sources with calibrated Kron magnitudes in both NUV and FUV bands were selected. After cross-matched with the 
MIR sample with radius of 4$\arcsec$, $421$ galaxies were left. We need to use the formula from \citet{treyer07} to compute the 
intrinsic extinction of the sources, but this formula is not suitable for the red sequence galaxies, which were defined with the 
$NUV-r$ color $<$ 4 (at redshift 0.1). Finally, $359$ galaxies were included in our UV sample. The calibration accuracy of the 
measurements is about 3\% \footnote{http://galexgi.gsfc.nasa.gov/docs/galex/Cycle3/}.

\subsection{Spectral Classification}
\begin{figure}
\figurenum{1}
\epsscale{}
\plotone{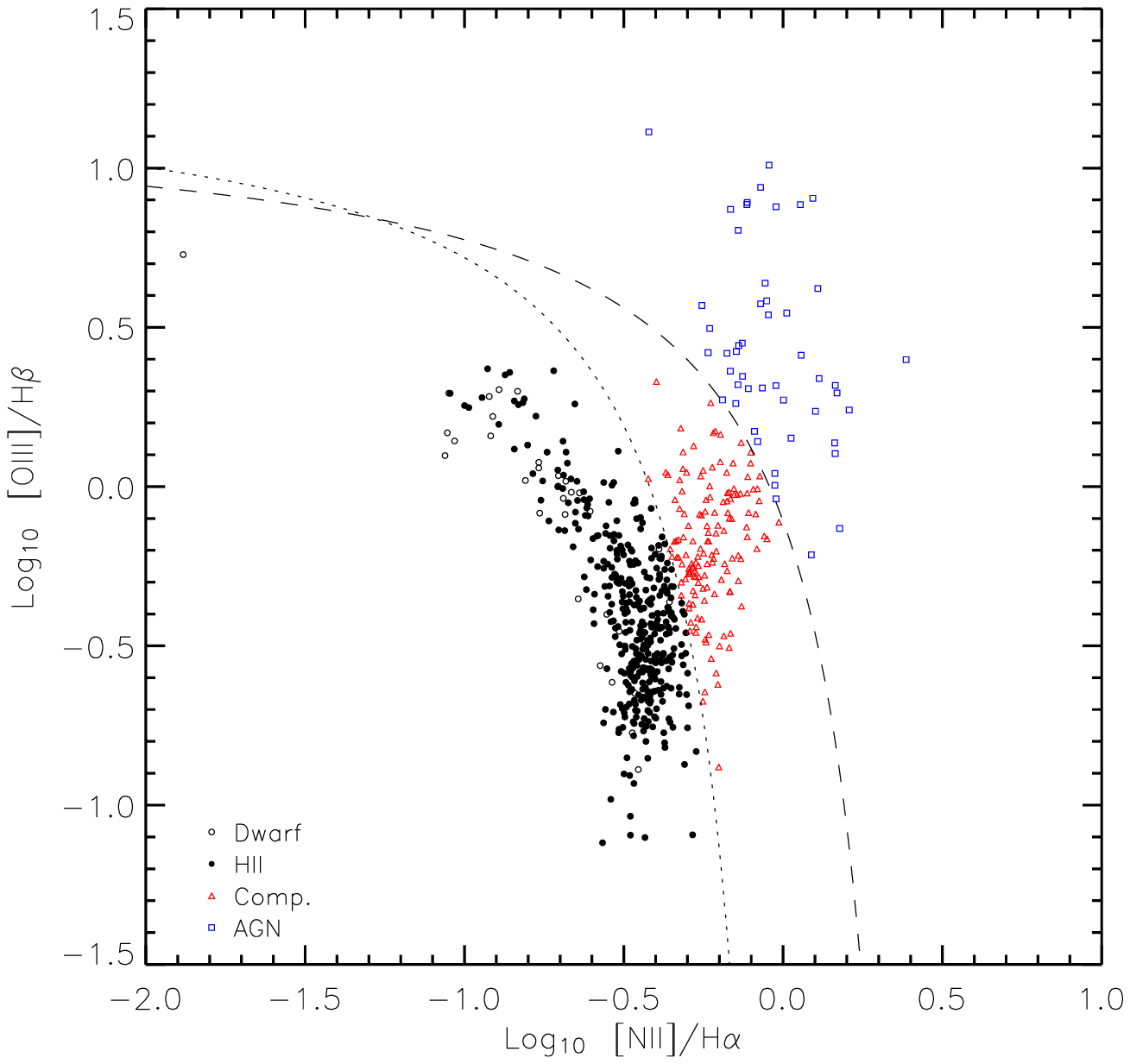}
\caption{The BPT diagnostic diagram: [NII]/H$\alpha$ vs [OIII]/H$\beta$. The criteria from \citet{kauffmann03} and \citet{kewley01} are 
illustrated as dotted and dashed curves, respectively. The objects below the dotted curve are defined as star-forming galaxies, and the 
open circles represent dwarf galaxies with M$_{\rm B}$ $>$ -18 mag. The boxes above the dashed curve denote AGNs, while the triangles 
between the above two curves are those classified as composite galaxies.}
\label{fig1}
\end{figure}

The optical spectral classifications of galaxies were made by adopting the traditional BPT diagnostic diagram: [NII]/H$\alpha$ 
versus [OIII]/H$\beta$ \citep{baldwin81, veilleux87}, as shown in Figure~\ref{fig1}. The dashed curve is from \citet{kauffmann03} 
(hereafter Ka03 line) and the dotted curve is from \citet{kewley01} (hereafter Ke01 line). The objects located below the Ka03 line 
were classified as star-forming galaxies; those above the Ke01 line were classified as narrow-line AGNs (note we did not 
distinguish Seyfert~2s from LINERs \citep{heckman80} here); those between the above two lines were classified as composite 
(starburst $+$ AGN) galaxies \citep{kewley06}. Composite galaxies are similar to mixture-type galaxies defined by \citet{wu98}, 
which were based on three diagnostic diagrams (adding [SII]/H$\alpha$ vs. [OIII]/H$\beta$ and [OI]/H$\alpha$ vs. [OIII]/H$\beta$) 
from \citet{veilleux87}: if the classification of a source in the three diagnostic diagrams is different (e.g., it was shown in 
one diagram as a star-forming galaxy, but could be classified as a AGN by the others), then it was categorized as a mixture-type 
galaxy \citep{wu98}. The numbers on spectral classifications of our MIR, FIR and UV samples are shown in Table~\ref{tab2}.

\subsection{Sample Distribution}
\begin{figure}
\figurenum{2}
\epsscale{}
\plotone{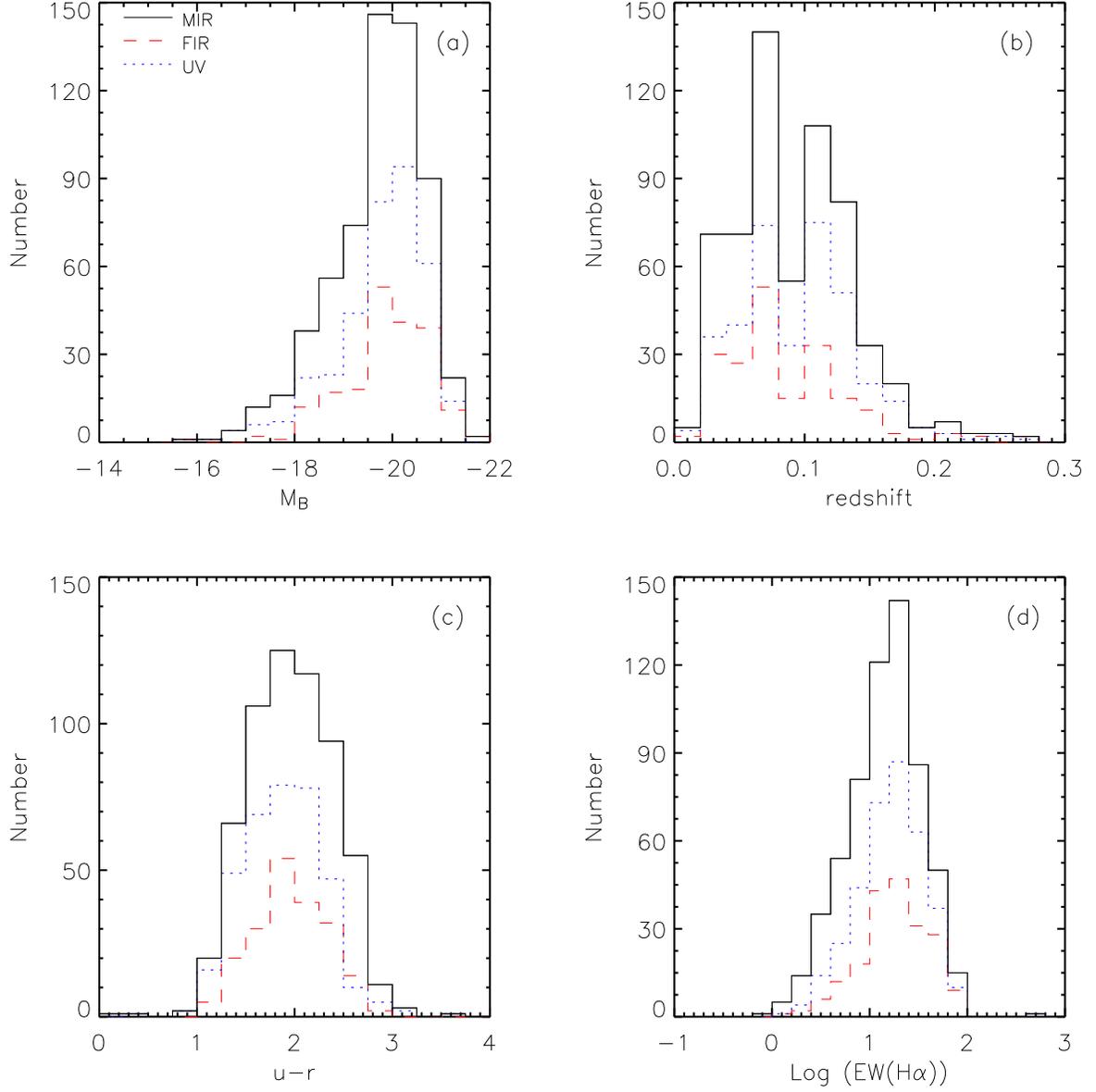}
\caption{The distributions of (a) absolute B-band magnitude; (b) redshift; (c) $u-r$ color; (d) EW(H$\alpha$) for galaxies in our MIR 
(black solid lines), FIR (red dashed lines), and UV (blue dotted lines) samples. The galaxy numbers of the above three samples are $605$, 
$197$ and $421$, respectively.}
\label{fig2}
\end{figure}

The distributions of the SDSS B-band absolute magnitude (M$_{\rm B}$), redshift, $u-r$ color, and equivalent width of H$\alpha$ line 
emission (EW[H$\alpha$]) for our MIR, FIR and UV sample galaxies are shown in Figure~\ref{fig2}. Most of the galaxies have M$_{\rm B}$ 
brighter than $-$18 mag and redshift less than 0.2. Only a few of them have magnitudes lower than $-$18 mag, which belong to dwarf 
galaxies \citep{thuan81}. Here, M$_{\rm B}$ were calculated from the SDSS $g$ and $r$-band magnitudes according to \citet{smith02}. 
The three galaxy samples do not show obvious differences in all of the above distributions.

Figure~\ref{fig3} shows the above four distributions for three different spectral type galaxies (star-forming galaxies, composite 
galaxies, and AGNs). No obvious difference was found in the distributions of either M$_{\rm B}$ or redshift. However, the distributions 
of $u-r$ color and EW(H$\alpha$) for composite galaxies and AGNs are strikingly different from those for star-forming galaxies. 
This result could be due to the fact that AGNs tend to reside in earlier type galaxies (\citealt{kauffmann03}; see $\S$5.4 for 
detailed discussions).

\section{Estimation of Luminosities}

\subsection{MIR luminosities}
To obtain MIR dust emission luminosities, the MIR continuum contributed by star light should be firstly estimated and then removed. 
A factor of 0.232 \citep{helou04} was used to scale the stellar continuum of 3.6$\mu$m to that of 8$\mu$m, with the assumption that 
the entire IRAC 3.6$\mu$m band emission is from stellar emission, and based on the {\sl Starburst99} synthesis model \citep{leitherer99}, 
assuming solar metallicity and a Salpeter initial mass function (IMF) between 0.1 and 120 $M_{\rm \odot}$. The scaling factor used 
here is not quite different from that of 0.26 by Wu05, which was obtained based on statistics of elliptical galaxies in the FLS field. 
Following Wu05, we denoted 8$\mu$m(dust) as the 8$\mu$m dust emission. The star light contribution to the MIPS 24$\mu$m band can be 
neglected since it decreases quickly with the increase of wavelength. 

Since IRAC 8$\mu$m band cover the complex PAH emission range, a more accurate spectral energy distribution(SED) model is needed to 
obtain a reliable MIR K-correction. Here we adopted a series of SEDs from \citet{huang07}'s two-component model which is a linear 
combination of an old 'early-type' stellar population and a 'late-type' spiral disk population. The MIR color [5.8]-[8.0] was used 
to select the best SED model, and then the K-correction value was obtained. A template SED of a normal HII galaxy NGC~3351 (from SINGS; 
\citealt{kennicutt03}) was used to perform K-correction for the 24$\mu$m band for all the sample galaxies. 

\subsection{FIR and TIR luminosities}
The K-correction of FIR 70 and 160$\mu$m emissions was based on the best fitted SED model of starburst galaxy M82 \citep{siebenmorgen07}. 
The correction factors are less than 12\% and 30\% for the above two bands. The TIR luminosity (3-1100$\mu$m) was calculated using the 
Equation (4) of \citet{dale02}:
\begin{equation}
L{[TIR]}=1.559\times\nu\,L_{\nu}{[24\mu\,m]}+0.7686\times\nu\,L_{\nu}{[70\mu\,m]}+1.347\times\nu\,L_{\nu}{[160\mu\,m]}
\end{equation}
%

\subsection{H$\alpha$ luminosity}
The optical spectra of SDSS were taken with 3$\arcsec$ diameter fibers, thus the measured H$\alpha$ emission fluxes were only from 
the central regions for most low redshift galaxies and the corresponding aperture correction is needed. We adopted the method by 
\citet{hopkins03} and used their Equation (A2) to obtain the H$\alpha$ luminosity of the whole galaxy:
\begin{equation}
L{[H\alpha]} = 4\pi D_{\rm L}^{2}\,S_{\rm H\alpha}\,10^{-0.4(r_{\rm Petro} - r_{\rm fiber})},
\end{equation}
Here, $D$$_{\rm L}$ is the luminosity distance; $S$$_{\rm H\alpha}$ is flux of H$\alpha$ emission; while $r$$_{\rm Petro}$ and $r$$_{\rm fiber}$ 
are the $r$-band Petrosian and fiber magnitudes, respectively. To match the 6$\arcsec$ aperture photometry in 24$\mu$m band, we also 
calculated the H$\alpha$ luminosities in aperture of 6$\arcsec$ in a similar way, i.e., substituting the $r$-band 6$\arcsec$ $aperture$ 
magnitudes for $r$$_{\rm Petro}$ in the above equation. The $r$-band 6$\arcsec$ $aperture$ magnitude was derived from the $r$-band radial 
surface brightness profile of the SDSS catalog \citep{stoughton02}.

The optical H$\alpha$ emissions suffer from the dust extinctions from both Milky Way and the host galaxy. The foreground Galactic 
extinction was first corrected by assuming the \citet{cardelli89}'s extinction curve and R$_{\rm V}$=3.1. Then the intrinsic extinction 
correction was performed, based on the color excess E(B-V) which was acquired from the Balmer decrement $F_{\rm H\alpha}/F_{\rm H\beta}$ 
\citep{calzetti01}.

\subsection{FUV luminosity}
The K-correction of UV band was based on the IDL code supplied by Blanton (version tag v4\_1\_4). The method and the SEDs used by 
this code were described by \citet{blanton03} and \citet{blanton07}. Since dust will heavily obscure the UV emission, it is important 
to correct for the UV extinction. The Galactic extinction was determined by the parametrized curve of \citet{cardelli89}, with the 
adopted conversion factors of 7.9 and 8.0 \citep{gil07} for FUV and NUV, respectively \citep{buat05}. The intrinsic UV extinctions 
are usually computed by the TIR-to-UV ratio method \citep{kong04, buat05, seibert05, cortese06, zamojski07, salim05, salim07}. However, 
the number of our UV sample galaxies with FIR data is too small to make statistical analysis. Alternatively, using a large sample of 
galaxies from SDSS spectroscopic catalog and GALEX Medium Imaging Survey, \citet{treyer07} derived an empirical correlation to estimate 
the intrinsic UV extinction based on the $FUV-NUV$ color at redshift $\sim$ 0.1 (their equation (9)):
\begin{equation}
A_{FUV} = 4.05\times(m_{FUV}-m_{NUV})-0.18 .
\end{equation}
We use this equation to calculate the intrinsic UV extinctions for our UV sample galaxies.

\section{Correlation Analysis}

\subsection{L(MIR) vs. L(H$\alpha$)}

{\noindent \large \it $Total$ luminosity}

Firstly, we present the correlation between $total$ MIR luminosities and $total$ H$\alpha$ luminosities for star-forming galaxies. 
The $total$ H$\alpha$ luminosities are the aperture-corrected values (see $\S$3.3). $Total$ 24$\mu$m and 8$\mu$m(dust) luminosities 
are plotted against $total$ H$\alpha$ luminosities in Figure~\ref{fig4} (a) and (b). After excluding the dwarf galaxies (open circles), 
star-forming galaxies show good correlations between $total$ L(MIR) and L(H$\alpha$). Using the two-variable regression, we obtain the 
best nonlinear fits ($\log_{10}(y)=$a$+$b$\log_{10}(x)$; solid lines in Figure~\ref{fig4} (a) and (b)):
\begin{equation}
Log_{10} { ~\nu L_{\nu}{[24\mu m]}_{total}}=(-0.06\pm0.07)+(1.18\pm0.02)\times Log_{10} ~L[H_{\rm \alpha}]_{total},
\end{equation}
\begin{equation}
Log_{10} { ~\nu L_{\nu}{[8.0\mu m(dust)]}_{total}} =(1.14\pm0.07)+(1.07\pm0.03)\times Log_{10} ~L[H_{\rm \alpha}]_{total}.
\end{equation}
Comparing with the nonlinear fits for star-forming galaxies in the $Spitzer$ FLS field obtained by Wu05 (dashed lines in Figure~\ref{fig4} 
(a) and (b)), we find that the best nonlinear fit of $total$ 24$\mu$m vs. H$\alpha$ luminosities in this work is well consistent with the 
corresponding result by Wu05. However, the slope of our nonlinear fit of $total$ 8$\mu$m(dust) vs. H$\alpha$ is a little shallower than 
that of Wu05, and this discrepancy could be due to the 8$\mu$m-weak HII galaxies (located at bottom-left of Figure~\ref{fig4} (b)) which 
tend to have lower 8$\mu$m(dust)-to-H$\alpha$ luminosity ratios.

Similar to Wu05, we can derive MIR SFR formulae based on the best linear fits ( $\log_{10}(y)=$c$+\log_{10}(x)$; dotted lines in 
Figure~\ref{fig4} (a) and (b), see also Table~\ref{tab3} for the fitting parameters) and Formula (2) of \citet{kennicutt98}:
\begin{equation}
SFR(M_{\odot} yr^{-1}) = 7.9\times 10^{-42}{L[H_{\rm \alpha}]} (erg s^{-1}),
\end{equation}
Here, solar abundances and the Salpeter IMF (0.1-100 $M_{\odot}$) were used in deriving the calibration factor. So the derived formulae 
according to the above correlations are:
\begin{equation}
SFR_{\rm 24\mu m}(M_{\odot} yr^{-1}) = \frac{\nu L_{\nu}[24\mu m]_{total}}{7.15\times 10^{8} L_{\odot}},
\end{equation}
\begin{equation}
SFR_{\rm 8\mu m(dust)}(M_{\odot} yr^{-1}) = \frac{\nu L_{\nu}[8\mu m(dust)]_{total}}{1.58\times 10^{9} L_{\odot}}.
\end{equation}
The fitting coefficients are consistent with those derived by Wu05 with differences of $\sim$11\%.

{\noindent \large \it $Aperture$ luminosity}

In the above analysis, the aperture effect was corrected and then the $total$ H$\alpha$ luminosities were obtained based on the 
assumption that the outer parts of galaxies have the same H$\alpha$-to-$r$ luminosity ratios to the central regions. In fact, this 
could not be true for many reasons, e.g., for central starburst galaxies, it would overestimate the $total$ H$\alpha$ fluxes. To 
evaluate such an effect, we re-analyzed above correlations only for the central regions of star-forming galaxies. Figure~\ref{fig4} 
(c) and (d) show the correlations between L(MIR) and L(H$\alpha$) in aperture of 6$\arcsec$ at 24~$\mu$m and of 3$\arcsec$ at 8$\mu$m. 
The best nonlinear fits are:
\begin{equation}
Log_{10} { ~\nu L_{\nu}{[24\mu m]}_{aper}}=(0.63\pm0.05)+(1.10\pm0.02)\times Log_{10} ~L[H_{\rm \alpha}]_{aper},
\end{equation}
\begin{equation}
Log_{10} { ~\nu L_{\nu}{[8.0\mu m(dust)]}_{aper}}=(1.74\pm0.05)+(1.02\pm0.02)\times Log_{10} ~L[H_{\rm \alpha}]_{aper}.
\end{equation}
Obviously, the scatters shown in Panels (c) and (d) (the standard deviation of the fitting residuals are 0.15 and 0.14 respectively) 
are smaller than those in Panels (a) and (b) (the standard deviation of the fitting residuals are 0.16 and 0.18 respectively) of 
Figure~\ref{fig4}, especially for the correlation between 8$\mu$m(dust) vs. H$\alpha$ luminosities (see also the fitting coefficients 
summarized in Table~\ref{tab3}). Compared with the best nonlinear fits for $total$ L(MIR) vs. L(H$\alpha$) (dashed lines in 
Figure~\ref{fig4} (c) and (d)), the best nonlinear fits for $aperture$ L(MIR) vs. L(H$\alpha$) (solid lines in Figure~\ref{fig4} 
(c) and (d)) are about 0.1~dex shifted upward. This result indicates that the MIR-to-H$\alpha$ luminosity ratios of the central 
regions are relatively higher than those of the whole galaxies. Panels (c) and (d) (``$aperture$'' correlations) also show that 
the deviations of dwarf galaxies from normal star-forming galaxies are much less than those shown in Panels (a) and (b) 
(``$total$'' correlations).

\subsection{L(H$\alpha$$_{\rm obs}$ $+$ MIR) vs. L(H$\alpha$)}
\citet{calzetti07} showed that there existed a tight correlation between the linear combination of the observed H$\alpha$ 
(H$\alpha$$_{\rm obs}$) and 24$\mu$m luminosities and the extinction-corrected Pa$\alpha$ luminosities for star-forming knots 
in SINGS galaxies. This result can be interpreted as: the 24$\mu$m (and also 8$\mu$m[dust]) and observed H$\alpha$ emissions 
trace the dust-obscured and un-obscured star formation, respectively. This correlation could be presented as equation 
illustrated by \citet{kennicutt07} (their equation 4): 
\begin{equation}
L[H_{\alpha}]=L[H_{\alpha\_obs}]+\alpha \times \nu L_{\nu}{[24\mu m]}.
\end{equation}
Here, the left side of this equation is the extinction-corrected H$\alpha$ luminosities.

Based on these star-forming galaxies, we can obtain the scaling constant $\alpha$ \citep{kennicutt07} values of 0.022 and 0.010 for the 
combined H$\alpha$$_{\rm obs}$ $+$ 24$\mu$m and H$\alpha$$_{\rm obs}$ $+$ 8$\mu$m(dust) luminosities. The $\alpha$ value of 0.022 we 
obtained here is slightly smaller than that of 0.031 for star-forming regions in galaxies by \citet{calzetti07}, and we speculate that 
this difference is caused by the dust heated by the interstellar radiation field (mainly from old stars) which could enhance the MIR 
emissions for the entire galaxies. In addition, the tight correlations between the dust-obscured (absorbed) H$\alpha$ luminosities 
and both of the 24$\mu$m and 8$\mu$m(dust) luminosities also verify the fact that the dominant sources of MIR emissions are from the 
dust-obscured star formation because the H$\alpha$ emission directly traces the HII region surrounded young massive stellar population. 
Therefore, the linear combinations of the H$\alpha$$_{\rm obs}$ and MIR luminosities would include almost all the star formation activities 
in galaxies, and they could be used as good SFR indicators for normal star-forming galaxies.

With the derived constant $\alpha$, the tight correlations between combined H$\alpha$$_{\rm obs}$ $+$ MIR and extinction-corrected H$\alpha$ 
luminosities (see Table~\ref{tab3}) are shown in Figure~\ref{fig5}. The dwarf galaxies also follow the correlations for normal star-forming 
galaxies. The best nonlinear fits are shown as:
\begin{equation}
Log_{10} ({L[H_{\alpha\_obs}] + 0.022~\nu L_{\nu}{[24\mu m]}})=(0.29\pm0.04)+(0.96\pm0.01)\times Log_{10} ~L[H_{\alpha}].
\end{equation}
\begin{equation}
Log_{10} ({L[H_{\alpha\_obs}] + 0.010~\nu L_{\nu}{[8\mu m]}})=(0.86\pm0.04)+(0.89\pm0.01)\times Log_{10} ~L[H_{\alpha}].
\end{equation}
Thus, the combined L(H$\alpha$$_{\rm obs}$ $+$ 24$\mu$m) almost linearly correlate with extinction-corrected H$\alpha$ luminosities.

\subsection{L(MIR) vs. L(FIR)}
FIR luminosity was thought to be a good SFR tracer by previous $IRAS$ and $ISO$ observations. Figure~\ref{fig6} shows the correlations 
between MIR (24$\mu$m and 8$\mu$m[dust]) and FIR (70$\mu$m and 160$\mu$m) luminosities. Due to the relatively lower sensitivities of 
$Spitzer$ MIPS 70 and 160$\mu$m observations, only $141$ star-forming galaxies have both 70$\mu$m and 160$\mu$m detections. The best 
nonlinear and linear fits are shown as the solid and dotted lines in Figure~\ref{fig6}, and the fitting parameters are listed in 
Table~\ref{tab3}. Among the above four correlations, the L(24$\mu$m)-L(70$\mu$m) (Figure~\ref{fig6}(a)) and L(8$\mu$m[dust])-L(160$\mu$m) 
(Figure~\ref{fig6}(d)) are shown to be much tighter than the rest two correlations (Figure~\ref{fig6}(b) and (c)). The former two correlations 
have scatters with standard deviations of $\sim$0.07, while those of the latter two are about 0.10. Furthermore, from Table~\ref{tab3}, it 
is clear that the slopes of best nonlinear fits of L(24$\mu$m)-L(70$\mu$m) and L(8$\mu$m[dust])-L(160$\mu$m) correlations are both $\sim$1.04, 
which indicate that these two correlations are almost linear.

\subsection{L(MIR) \& L(FIR) vs. L(TIR)}
The correlations between MIR \& FIR with TIR (3-1100$\mu$m) luminosities are shown in Figure~\ref{fig7}. The fitting parameters and 
correlation coefficients are listed in Table~\ref{tab3}. All of the four correlations are tight, with the standard deviations of 0.06 
to 0.09 and the Spearman Rank-order correlation analysis coefficients of $\sim$0.97. The L(160$\mu$m)-L(TIR) correlation is the tightest 
among the four, and the 8$\mu$m(dust) and 160$\mu$m luminosities are almost linearly correlated with the TIR luminosities.

Based on these correlations, we re-calibrated the MIR SFR formulae from TIR luminosities. \citet{kennicutt98} presented the equation 
(Equation (4)) to calculate SFR based on IR luminosity:
\begin{equation}
SFR(M_{\odot} yr^{-1}) = 4.5\times 10^{-44}{L{[FIR]}} (erg s^{-1}),
\end{equation}
The calibration factor was derived by applying the models of \citet{leitherer95} for continuous bursts of age 10-100 Myr and adopting the 
same Salpeter IMF. Here, $L_{FIR}$ refers to the IR luminosity integrated over the full-, mid-, and far-IR spectrum (8-1000 $\mu$m). The 
TIR (3-1100$\mu$m) luminosity can also be derived from $IRAS$ three bands (25,60,100$\mu$m) luminosities \citep{dale02}, and 
\citet{takeuchi05} found that the L(TIR) estimated in this way were very similar to the L(IR) (8-1000$\mu$m; calculated by using the 
equations presented by \citealt{sanders96}). Therefore, we adopted above Equation and used the TIR luminosities to calibrate SFRs. 
The derived formulae are:
\begin{equation}
SFR_{\rm 24\mu m}(M_{\odot} yr^{-1}) = \frac{\nu L_{\nu}[24\mu m]}{5.62\times 10^{8} L_{\odot}},
\end{equation}
\begin{equation}
SFR_{\rm 8\mu m(dust)}(M_{\odot} yr^{-1}) = \frac{\nu L_{\nu}[8\mu m(dust)]}{1.17\times 10^{9} L_{\odot}}.
\end{equation}  
The fitting coefficients are different from those in Equations (6) and (7) by $\sim$ 20-25\%.

\citet{kennicutt98} pointed out that the calibration factor of 4.5 in their Equation (4) was not valid for quiescent, normal galaxies, 
since the lower optical depth of dust could increase this coefficient, while the dust heating from old stars could decrease it. The 
old stellar population could heat the interstellar dust and be re-radiated at longer wavelength as cool infrared cirrus emissions 
\citep[see, e.g.,][]{beichman87, choi06}. Therefore, the SFR derived from TIR luminosities would be overestimated if such contributions 
to L(TIR) can not be neglected. \citet{bell03} illustrated that the fraction of contribution to L(TIR) from diffuse components was 
$\sim$32$\%$ for normal star-forming galaxies, while this proportion is much smaller for Luminous Infrared Galaxies (LIRGs) with 
L(TIR) $> 10^{11}L_{\odot}$. This interpretation could explain the differences between fitting coefficients of SFRs calibrated by 
TIR and H$\alpha$ luminosities.

Considering that most of our sample galaxies have infrared luminosities less than 10$^{11}$ L$_{\odot}$, we adopt an old stellar 
population fraction of $\sim$30\% independent of luminosity \citep{hopkins03}. Then we obtain old stellar population corrected 
SFR formulae for star-forming galaxies (with L(TIR) $< 10^{11}L_{\odot}$):
\begin{equation}
SFR_{\rm 24\mu m}(M_{\odot} yr^{-1}) = \frac{\nu L_{\nu}[24\mu m]}{7.79\times 10^{8} L_{\odot}},
\end{equation}
\begin{equation}
SFR_{\rm 8\mu m(dust)}(M_{\odot} yr^{-1}) = \frac{\nu L_{\nu}[8\mu m(dust)]}{1.70\times 10^{9} L_{\odot}}.
\end{equation}  
The fitting coefficients are consistent with those in Equations (6) and (7), with much smaller differences of $\sim$8\%.

\subsection{L(MIR) vs. L(FUV)}
Figure~\ref{fig8} shows the 24$\mu$m and 8$\mu$m(dust) vs. extinction-corrected FUV luminosities (see $\S$3.4) for $248$ star-forming 
galaxies. Although scatters are large (with standard deviations of $\sim$0.22), there still exist correlations between L(MIR) and 
L(FUV) with the Spearman Rank-order correlation analysis coefficients of $\sim$0.86, 0.84, respectively. The conversion 
between the UV luminosity and SFR has been presented by \citet{kennicutt98} (Equation (1)):
\begin{equation}
SFR(M_{\odot} yr^{-1}) = 1.4\times 10^{-28}{L_{\nu}} (erg s^{-1} Hz^{-1}),
\end{equation}
The calibration factor was derived by the assumption of continuous star formation and adopting the same Salpeter IMF. Based on this 
formula, we derive SFRs from the FUV luminosities and the linear correlation coefficients listed in Table~\ref{tab3}:
\begin{equation}
SFR_{\rm 24\mu m}(M_{\odot} yr^{-1}) = \frac{\nu L_{\nu}[24\mu m]}{4.48\times 10^{8} L_{\odot}},
\end{equation}
\begin{equation}
SFR_{\rm 8\mu m(dust)}(M_{\odot} yr^{-1}) = \frac{\nu L_{\nu}[8\mu m(dust)]}{0.93\times 10^{9} L_{\odot}}.
\end{equation}  
The fitting coefficients are nearly half of those in Equations (6) and (7), but note the FUV luminosities only cover a range 
of $\sim$2 orders of magnitude.

However, \citet{kennicutt98}'s formula is only suit for continuous star formation over time scales of $10^{8}$ year or longer. 
\citet{burgarella05} presented a formula (their Equation (2)) to estimate SFRs for non-starburst galaxies whose stellar 
birthrate $b$ \citep{kennicutt94} are less than 1.0. Based on the relationship between $b$ and EW(H$\alpha$) given by 
\citet{kennicutt94}, we selected 187 star-forming galaxies whose EW(H$\alpha$) are less than 30 (correspond to $b < 1$) 
to re-calibrate SFRs:
\begin{equation}
SFR_{\rm 24\mu m}(M_{\odot} yr^{-1}) = \frac{\nu L_{\nu}[24\mu m]}{7.04\times 10^{8} L_{\odot}},
\end{equation}
\begin{equation}
SFR_{\rm 8\mu m(dust)}(M_{\odot} yr^{-1}) = \frac{\nu L_{\nu}[8\mu m(dust)]}{1.46\times 10^{9} L_{\odot}}.
\end{equation}  

Similar to the analysis in $\S$4.2, Figure~\ref{fig9} presents the correlations between the extinction-corrected FUV luminosities 
and combined observed FUV + MIR luminosities (L[FUV$_{\rm obs}$ $+$ MIR]), with derived coefficients $\alpha$ of 6.31 and 3.02 for 
FUV $+$ 24$\mu$m and 8$\mu$m, respectively. But the scatters of these correlations ($\sim$1.65) are larger than that of L(H$\alpha$) 
vs. L(H$\alpha$$_{\rm obs}$ $+$ MIR).

\subsection{Correlations for Composite galaxies and AGNs}
Besides the star-forming galaxies, there are also many composite galaxies and AGNs in our sample. The correlations between the 
$total$ \& $aperture$ MIR and H$\alpha$ luminosities for composite galaxies and AGNs are shown in Figure~\ref{fig10}. The parameters 
of the best nonlinear and linear fits for composite galaxies are listed in Table~\ref{tab4}. The L(MIR) are correlated with L(H$\alpha$) 
for both composite galaxies and AGNs, but the scatters are larger (with standard deviations of 0.15-0.21) and the correlation slopes 
are shallower than those of star-forming galaxies. Furthermore, the MIR-to-H$\alpha$ luminosity ratios of both composite galaxies 
and AGNs are higher than that of star-forming galaxies, especially for the L(24$\mu$m)-to-L(H$\alpha$) ratios. Similarly, both 
composite galaxies and AGNs are also shown to have relatively higher MIR-to-FUV luminosity ratios (see Figure~\ref{fig8}).

Figures~\ref{fig11} and~\ref{fig12} show the correlations of L(MIR) vs. L(FIR) and L(MIR) \& L(FIR) vs. L(TIR) for composite galaxies 
and AGNs. The best nonlinear and linear fits for composite galaxies are similar to those for star-forming galaxies but with larger 
scatters. Just the same as normal star-forming galaxies, for composite galaxies the L(24$\mu$m)-L(70$\mu$m) and L(8$\mu$m[dust])-L(160$\mu$m) 
correlations are tighter than those of L(24$\mu$m)-L(160$\mu$m) and L(8$\mu$m[dust])-L(70$\mu$m); the best nonlinear fits of the former 
two correlations are almost linear; and both 8$\mu$m(dust) and 160$\mu$m luminosities are almost linearly correlated with TIR luminosities.

\section{Discussion}

\subsection{Discrepancies in L(MIR)-L(H$\alpha$) Correlations}
The aperture correction to H$\alpha$ luminosities (see $\S$3.3) were performed based on the assumption that the distributions 
of H$\alpha$ emission in galaxies are similar to those of the continuum (SDSS-$r$) emission. All of our sample galaxies have both 
detectable H$\alpha$ and H$\beta$ emission lines above 5$\sigma$ significance level, thus it could bias towards galaxies with 
stronger (circum-)nuclear star formation activities. In the outer parts of these types of galaxies, star formation activity 
would be much weaker and the optical continuum emissions could be dominated by old stellar populations. In such cases, we would 
overestimate the $total$ H$\alpha$ luminosity if performing corrections from SDSS $r$-band magnitude. The aperture correction 
effect discussed above could explain the shifts between best fits of $total$ and $aperture$ MIR-to-H$\alpha$ luminosity 
correlations shown in Figures~\ref{fig4} and described in $\S$4.1.

Besides the aperture correction effect discussed above, differential dust obscuration could be another reason for such discrepancies. 
The central region of galaxies is more dusty than their outer parts \citep[e.g.,][]{popescu05,prescott07}, and a majority of UV continuum 
and Balmer line emissions might be completely enshrouded by dust thus of course can't be corrected. There is no doubt that the above 
effect will be more severe for the central regions than for the entire galaxies, thus it could cause a higher $aperture$ MIR-to-H$\alpha$ 
luminosity ratio than the $total$ value, in accordance with the discrepancies shown above.

Dwarf galaxies were shown to be obviously deviated from the star-forming galaxies in the L(MIR)-L(H$\alpha$) correlations (see 
Figure~\ref{fig4}(a) and (b)). This large deviation could be mainly caused by the aperture correction effect discussed above, since 
many dwarf galaxies have intensive star formation in their nuclear regions \citep[e.g.,][]{gu06,lisker06}. This interpretation 
can also be proved by the $aperture$ L(8$\mu$m[dust])-L(H$\alpha$) correlation shown in Figure~\ref{fig4}(d), in which dwarf 
galaxies are only slightly deviated from the star-forming galaxies. Alternatively, this slight deviation could be explained by 
the fact that PAH and warm dust emissions are relatively weak in low metallicity environments \citep[e.g.,][]{madden00, engelbracht05, 
wyl06, galliano05, wu07, engelbracht08}.

\subsection{Estimation of SFRs from MIR Luminosities}
Based on a sample of $Spitzer$ FLS-field galaxies, \citet{wu05} showed that both $Spitzer$ 8$\mu$m(dust) and 24$\mu$m luminosities 
could be used to calculate SFRs of normal star-forming galaxies, and they gave corresponding MIR-derived SFR formulae calibrated by 
either H$\alpha$ or radio luminosities. However, the size of Wu05's sample is a little small (only have $<$ 80 galaxies) and the 
MIR luminosities of their sample galaxies only cover a range of two orders of magnitude. In this work, we construct a large sample 
of $379$ star-forming galaxies with the MIR luminosities range of three orders of magnitude. Besides correlating well with H$\alpha$ 
luminosities, the MIR luminosities of our sample galaxies are also found to have tight correlations with TIR and FUV luminosities. 
These correlations allow us to estimate and compare SFRs from MIR luminosities calibrated by different SFR indicators.

Although the correlations between L(MIR) with L(H$\alpha$) \& L(TIR) are somewhat nonlinear, the nonlinearities of the above two 
correlations are generally less than 10\%. Therefore, we can adopt linear correlations for calibrating SFRs (SFRs derived from 
non-linear correlations between L(MIR) with other luminosities also were illustrated in Figure~\ref{fig13}) . Based on the three 
linear correlations of L(MIR)-L(H$\alpha$), L(MIR)-L(TIR) and L(MIR)-L(FUV), we obtained that the SFR conversion factors of L(24$\mu$m) 
are 7.15$\times$10$^8$, 7.79$\times$10$^8$, and 7.04$\times$10$^8$ L$_{\odot}$, and those of L(8$\mu$m[dust]) are 1.58$\times$10$^8$, 
1.70$\times$10$^8$, and 1.46$\times$10$^8$ L$_{\odot}$, respectively. These factors are in good agreement with those of L(MIR)-L(H$\alpha$) 
and L(MIR)-L(radio) correlations by Wu05 with differences less than 10\% (Figure~\ref{fig13}). In fact, the differences in L(8$\mu$m[dust])-L(H$\alpha$) 
correlations between Wu05's sample and ours are mainly because of the galaxies with very low MIR luminosities. However, it must 
be noted that the MIR SFR formulae (especially for 8$\mu$m(dust)) derived above can not be applied to dwarf galaxies (and other 
galaxies with low metallicity; \citealt{wu07}), since they deviate much from the linear correlations for normal star-forming galaxies. 
However, the deviation of dwarfs to normal star-forming galaxies in Figure~\ref{fig5} is apparently smaller than in Figure~\ref{fig4}. 
Therefore, it may be a simple and feasible method of using the combined observed H$\alpha$ + MIR luminosities to compute SFRs. 
The black solid line in Figure~\ref{fig13} show the equation to compute SFR based on 24$\mu$m luminosity given by \citet{alonso06a} 
(their equation 3), which was derived from the correlation between 24$\mu$m and H$\alpha$ luminosities for 30 LIRGs in the local 
universe, and this equation is similar to the one by \citet{calzetti07} except for different IMFs. We should keep in mind such 
large differences for infrared luminous galaxies.

For TIR emission, the contribution from dust heating by the old stellar population could not be ignored ($\sim$30\% for normal galaxies, 
\citet{bell03}); while for UV emission, except for the pollution from old star, the effect of intrinsic extinction is even stronger 
than that for H$\alpha$ emission. Although the standard deviations of the fitting residuals from TIR (0.07 for 24$\mu$m, 0.06 for 
8$\mu$m(dust)) are smaller than from H$\alpha$ (0.16, 0.18 respectively), the H$\alpha$ emission trace the HII region surrounded 
massive star more directly. Therefore, we suggest that the MIR-derived SFR based on the MIR vs. H$\alpha$ correlation is ``better'' 
than those derived from other correlations. 

\subsection{Correlations among L(MIR), L(FIR) and L(TIR)}
The MIR luminosities L(MIR) are mainly from PAH and warm dust (VSGs) emissions, while FIR luminosities L(FIR) are dominated by cold 
dust emissions from large grains. From Figure~\ref{fig6}, and comparing with the correlations between L(MIR) and L(H$\alpha$) shown 
in Figure~\ref{fig4} and Table~\ref{tab3}, we find that the MIR luminosities correlate better with L(FIR) than L(H$\alpha$), and this 
result is consistent with that from \citet{boselli04} for late-type galaxies observed by $ISO$.

However, correlations shown in Figures~\ref{fig4} and~\ref{fig6} are on the basis of two different (MIR \& FIR) samples. Therefore, 
we firstly checked the distributions of absolute B-band magnitude, redshift, $u-r$ color, and EW(H$\alpha$) for both MIR and FIR 
sample galaxies (see Figure~\ref{fig2}), and found the two samples have similar distributions. To avoid possible selection effect, 
the L(MIR)-L(H$\alpha$) correlations for FIR sample galaxies are also shown in Figure~\ref{fig14}. The best fits for galaxies in 
the FIR sample are consistent with those for Wu05 and MIR sample galaxies. The standard deviations for the FIR sample are about 0.15, 
a little smaller than those for the MIR sample, since the FIR sample do not include galaxies with L(MIR) $<$ 10$^8$$L_{\odot}$; while 
their Spearman Rank-order correlation analysis coefficients are similar to those of the MIR sample. From the comparison described 
above, we confirm that the MIR luminosities are better correlated with FIR than H$\alpha$ luminosities for star-forming galaxies.

As described in $\S$4.3 and shown in Figure~\ref{fig6}, among the four L(MIR)-L(FIR) correlations, L(24$\mu$m)-L(70$\mu$m) and 
L(8$\mu$m[dust])-L(160$\mu$m) have much smaller scatters (the standard deviation of the fitting residuals are both 0.07) than 
the rest two correlations of L(24$\mu$m)-L(160$\mu$m) and L(8$\mu$m[dust])-L(70$\mu$m) (the standard deviation of the fitting 
residuals are 0.12 and 0.10 respectively). Furthermore, the slopes of both L(24$\mu$m)-L(70$\mu$m) and L(8$\mu$m[dust])-L(160$\mu$m) 
correlations are close to unity, which indicates that such correlations are almost linear. The above results hint that the 24$\mu$m 
vs. 70$\mu$m, and 8$\mu$m vs. 160$\mu$m emissions may have similar physical origins:\\
$~~~~$(a) The 24$\mu$m emission is known to be dominated by warm dust emissions from VSGs, which are mainly heated by young massive stars. 
Recent results by $Spitzer$ \citep{wu05,perez06,calzetti07} have shown that the 24$\mu$m luminosity is one of the best SFR indicators. 
Previous works based on $IRAS$ and $ISO$ observations have proved that the FIR luminosity is also a good SFR tracer because of 
the emission peak around 60$\mu$m of dust heated by star formation, thus the 70$\mu$m emission must be closely related to star 
formation activities and should have tight and linear correlation with 24$\mu$m warm dust emission.\\
$~~~~$(b) The 8$\mu$m(dust) luminosity was also thought to be a SFR tracer by some authors \citep[e.g.,][]{wu05,alonso06b}. 8$\mu$m 
dust emissions include emissions from both PAHs and VSGs. In star-forming galaxies, the 7.7$\mu$m PAH emission dominates the 8$\mu$m 
band \citep{smith07} and its strength is likely to be affected by many factors, such as the radiation field and metallicity 
\citep[e.g.,][]{madden00,galliano05,cao07,wu07}. Moreover, PAH emissions trace B star better than young O stars \citep{peeters04}. 
Therefore, the 8$\mu$m(dust) luminosity may not be an optimal SFR tracer, and this could explain why the 70$\mu$m luminosities 
have better correlation with 24$\mu$m rather than 8$\mu$m(dust) luminosities.\\
$~~~~$(c) The 160$\mu$m emission is mainly from cold dust heated by old stars and is thought to locate in the diffuse regions 
of galaxies \citep[e.g.,][]{gordon04}; while the PAH molecules are generally located in the photo-dissociation regions (PDRs), 
at the interface of HII regions \citep{hollenbach97,draine03}. The tight L(8$\mu$m[dust])-L(160$\mu$m) correlation indicates 
that the 8$\mu$m(dust) and 160$\mu$m emissions might be from the same regions of galaxies, i.e., cold dust emission could 
exists in PDRs and it is also possible that there exists PAH emission in the diffuse regions because PAHs can be pumped by 
cool stars with less UV radiation \citep{li02}. Due to the low spatial resolution and shallowness of $Spitzer$ 160$\mu$m 
observations, firm conclusions on the physical connection between 8$\mu$m(dust) and 160$\mu$m emissions and their 
distributions in galaxies must await high quality FIR data from future space telescopes (e.g., $Herschel$).

Figure~\ref{fig6} shows that both MIR and FIR luminosities are well correlated with TIR luminosities for star-forming galaxies. 
Except for L(24$\mu$m), all of the 8$\mu$m(dust), 70$\mu$m and 160$\mu$m luminosities are almost linearly correlated with the 
L(TIR). This result indicates that the monochromatic MIR and FIR luminosities can be used for estimating the TIR luminosity, 
in agreement with previous studies \citep{elbaz02, takeuchi05, marcillac06}.

\subsection{Properties of Composite galaxies and AGNs}
\begin{figure}
\figurenum{3}
\epsscale{1}
\plotone{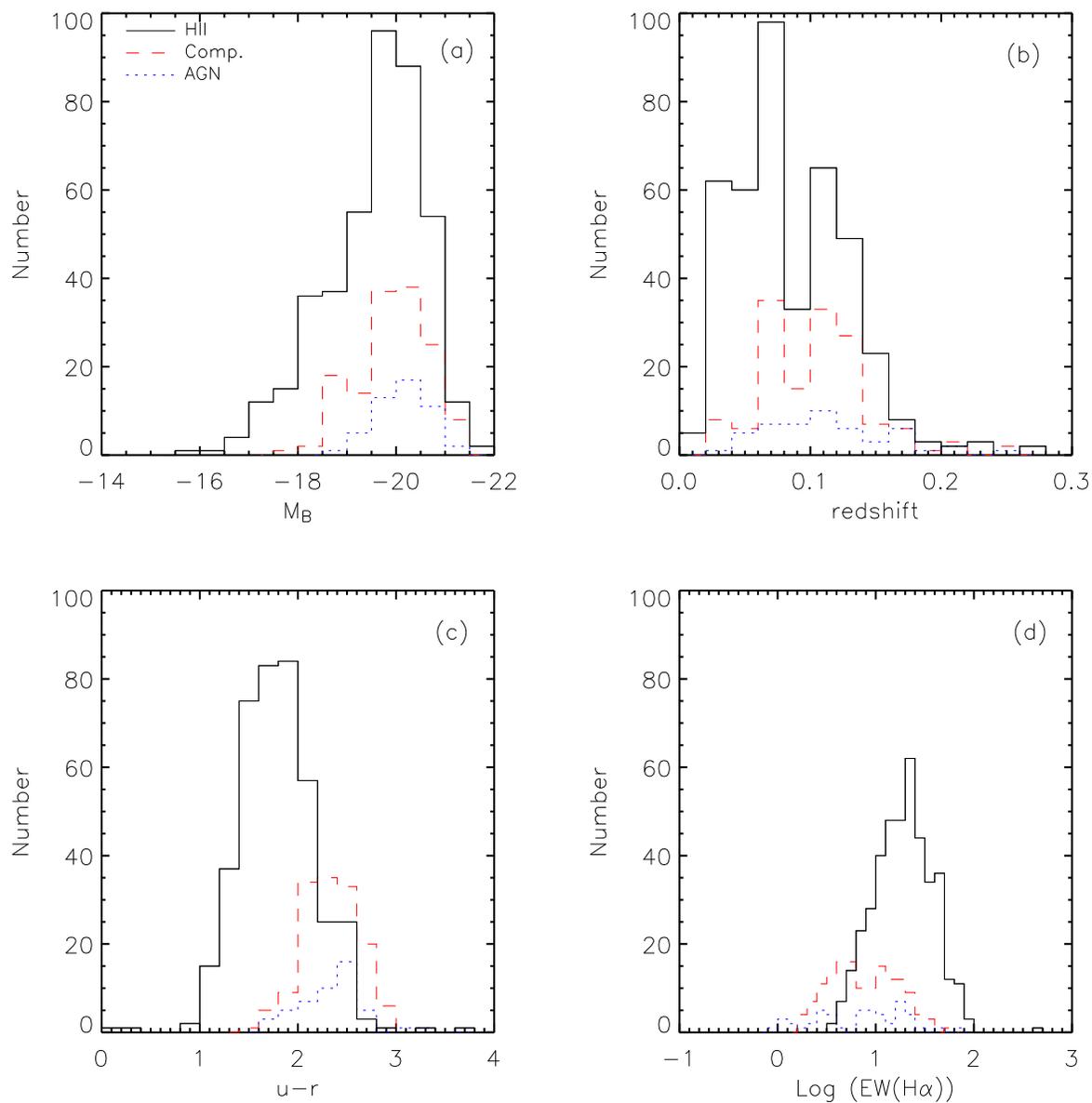}
\caption{The distributions of (a) absolute B-band magnitude; (b) redshift; (c) $u-r$ color; (d) EW(H$\alpha$) for $413$ star-forming 
galaxies (black solid lines), $143$ composite galaxies (red dashed lines), and $49$ AGNs (blue dotted lines) in our MIR sample.}
\label{fig3}
\end{figure}

\begin{figure}
\figurenum{4}
\epsscale{0.9}
\plotone{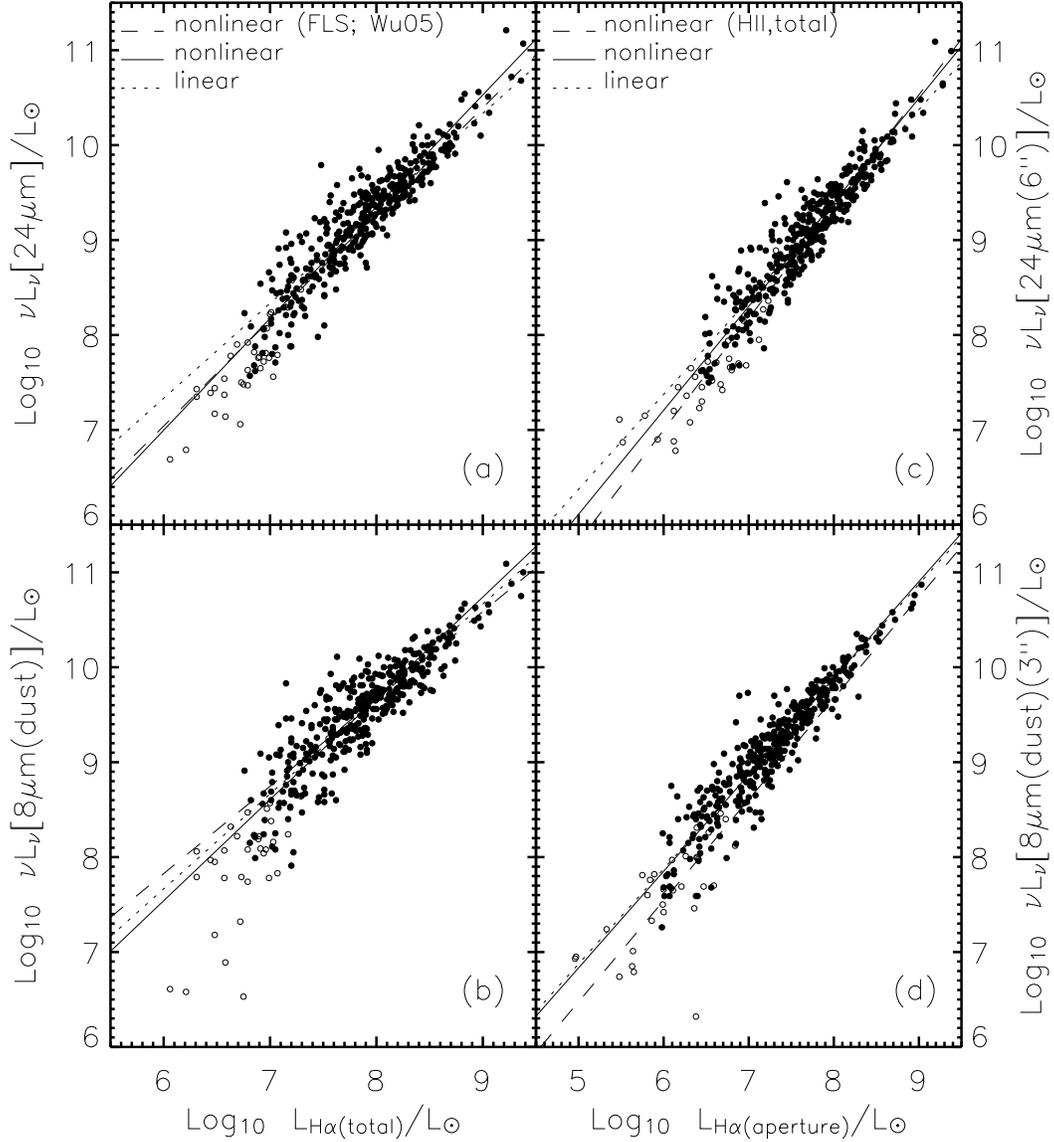}
\caption{Correlations between MIR and H$\alpha$ luminosities for star-forming galaxies. Panels (a) \& (b): $total$ H$\alpha$ vs. $total$ 
MIR luminosities; Panels (c) \& (d): $aperture$ H$\alpha$ vs. $aperture$ MIR luminosities. Dwarf galaxies are denoted as open circles. 
The best nonlinear and linear fits for normal star-forming galaxies (solid circles) are illustrated as solid and dotted lines. The 
dashed lines show in Panels (a) \& (b) represent the nonlinear fits for FLS-field galaxies by Wu05, and the dashed lines shown in Panels 
(c) \& (d) represent the nonlinear fits of $total$ MIR-to-H$\alpha$ luminosity correlations (solid lines in Panels (a) \& (b)).}
\label{fig4}
\end{figure}

\begin{figure}
\figurenum{5}
\epsscale{1}
\plotone{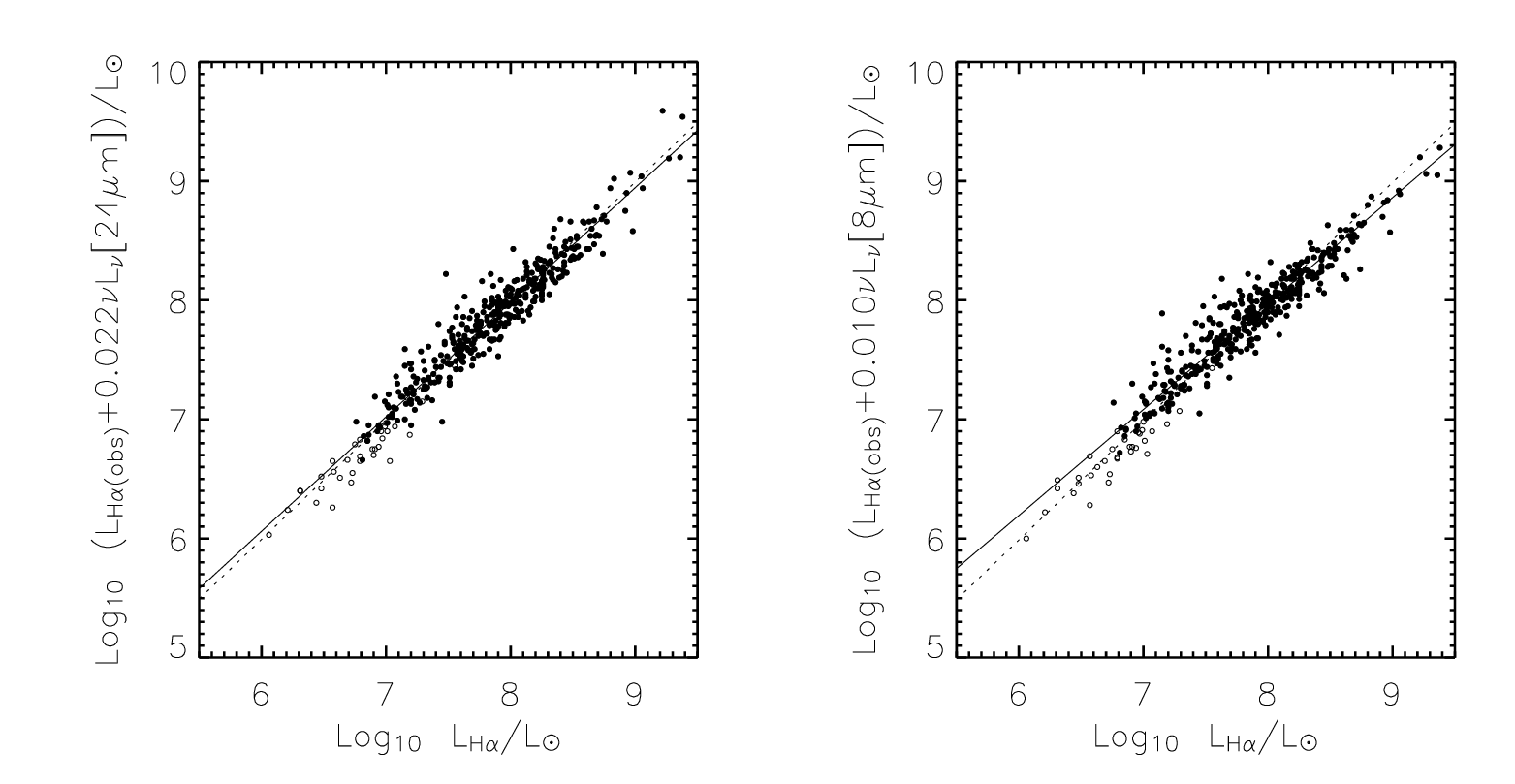}
\caption{Correlations between combined H$\alpha$$_{\rm obs}$ $+$ 24$\mu$m \& H$\alpha$$_{\rm obs}$ $+$ 8$\mu$m luminosities and 
extinction-corrected H$\alpha$ luminosities for star-forming galaxies. The symbols and line styles are the same as in Figure~\ref{fig4}.}
\label{fig5}
\end{figure}

\begin{figure}
\figurenum{6}
\epsscale{1}
\plotone{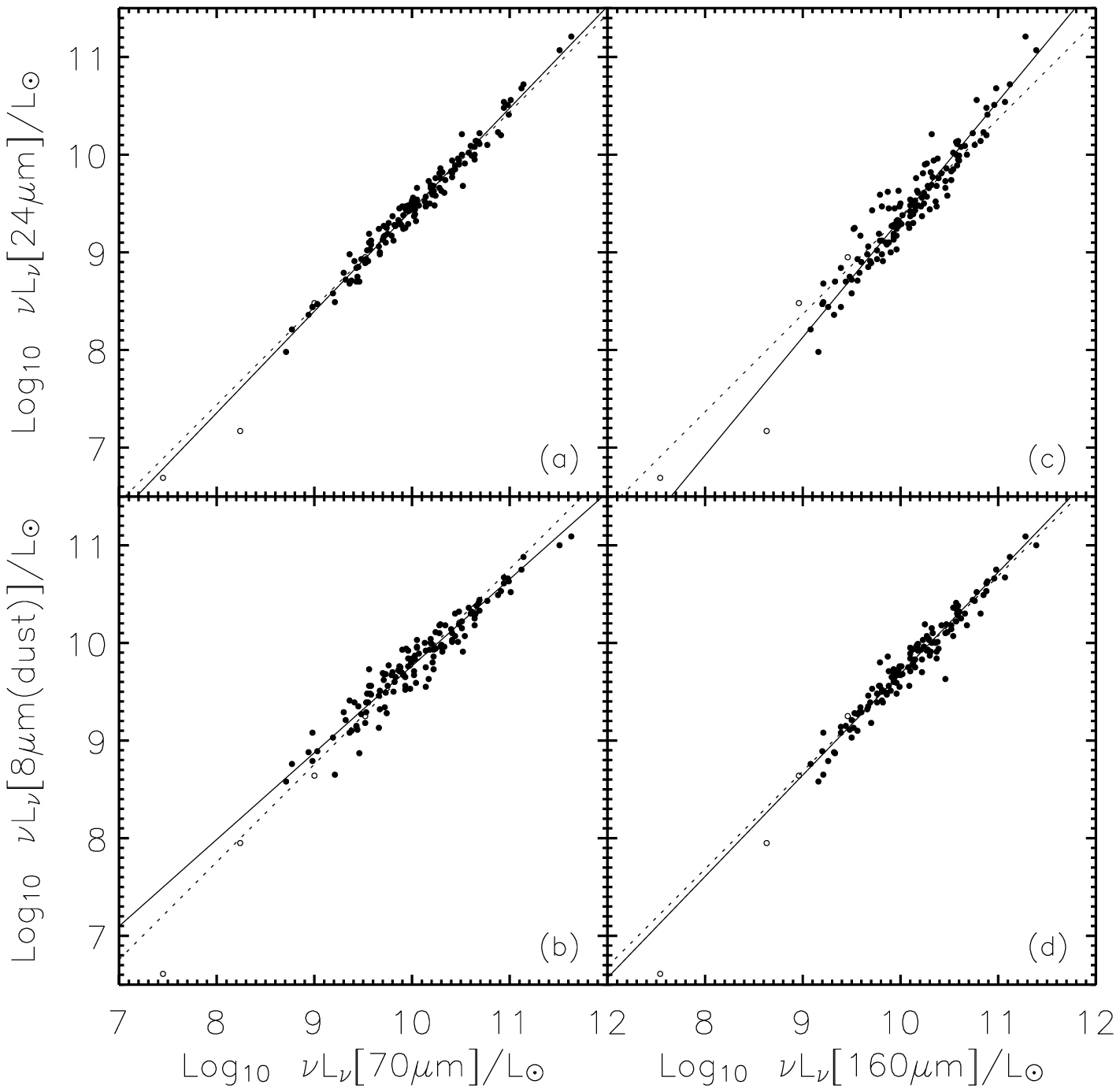}
\caption{Correlations between MIR and FIR luminosities for star-forming galaxies. The symbols and line styles are the same as in Figure~\ref{fig4}.}
\label{fig6}
\end{figure}

\begin{figure}
\figurenum{7}
\epsscale{1}
\plotone{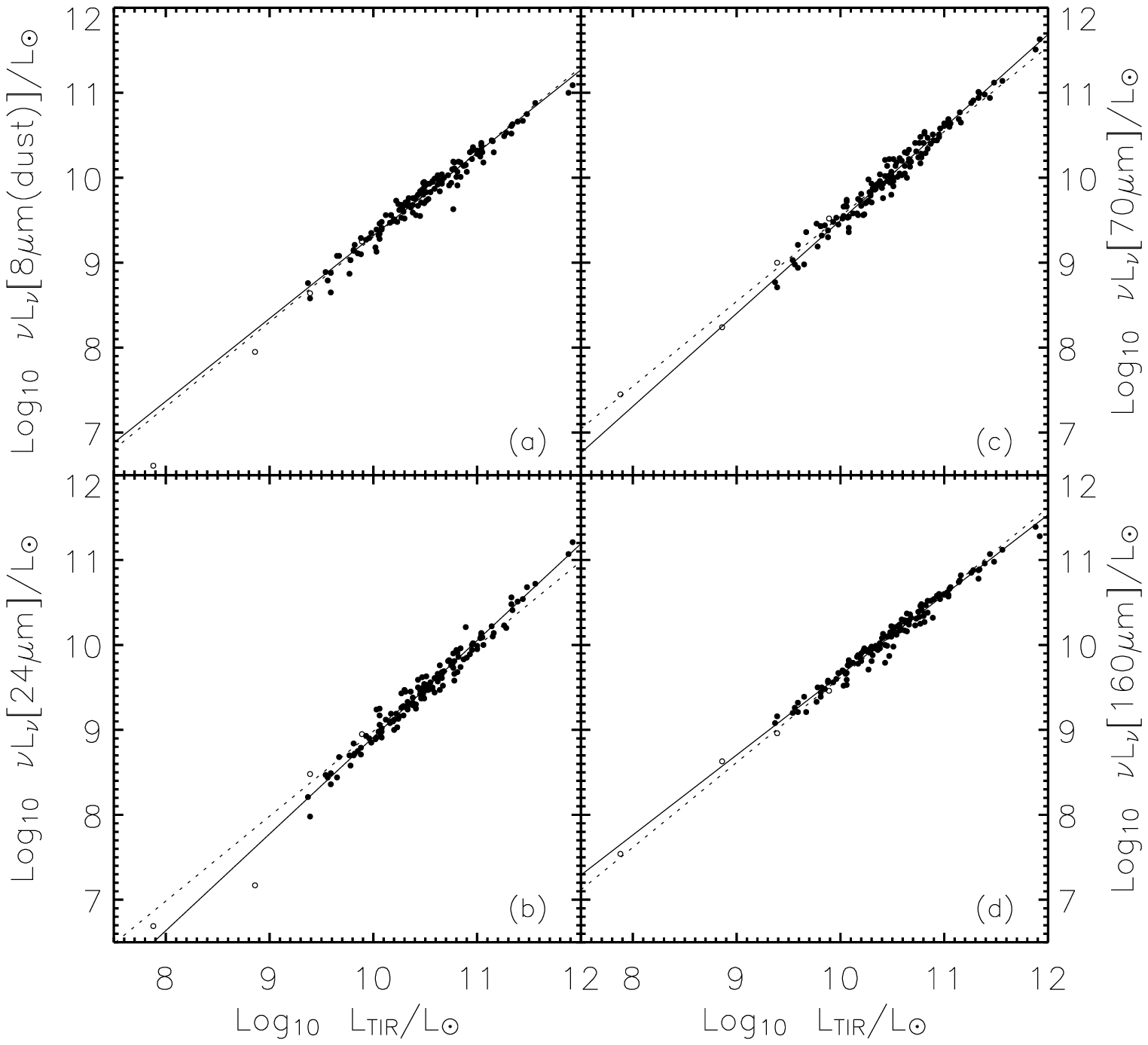}
\caption{Correlations between MIR \& FIR and TIR luminosities for star-forming galaxies. The symbols and line styles are the same as in Figure~\ref{fig4}.}
\label{fig7}
\end{figure}

\begin{figure}
\figurenum{8}
\epsscale{1}
\plotone{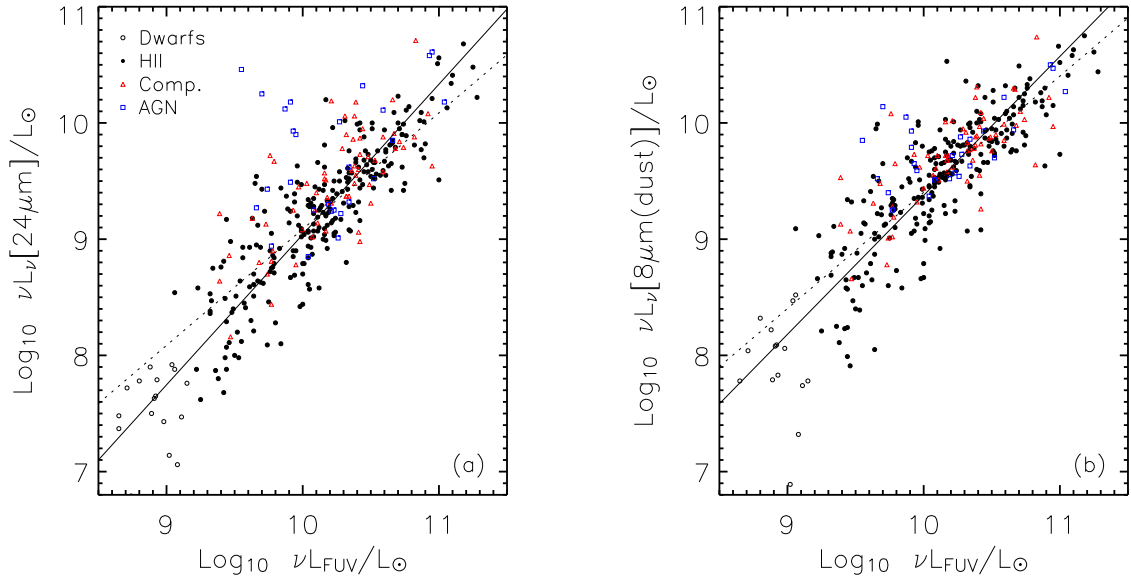}
\caption{Correlations between MIR and FUV luminosities for star-forming galaxies, composite galaxies and AGNs. The symbols are the same as in 
Figure~\ref{fig1}. The best nonlinear and linear fits for normal star-forming galaxies are illustrated as solid and dotted lines.}
\label{fig8}
\end{figure}

\begin{figure}
\figurenum{9}
\epsscale{1}
\plotone{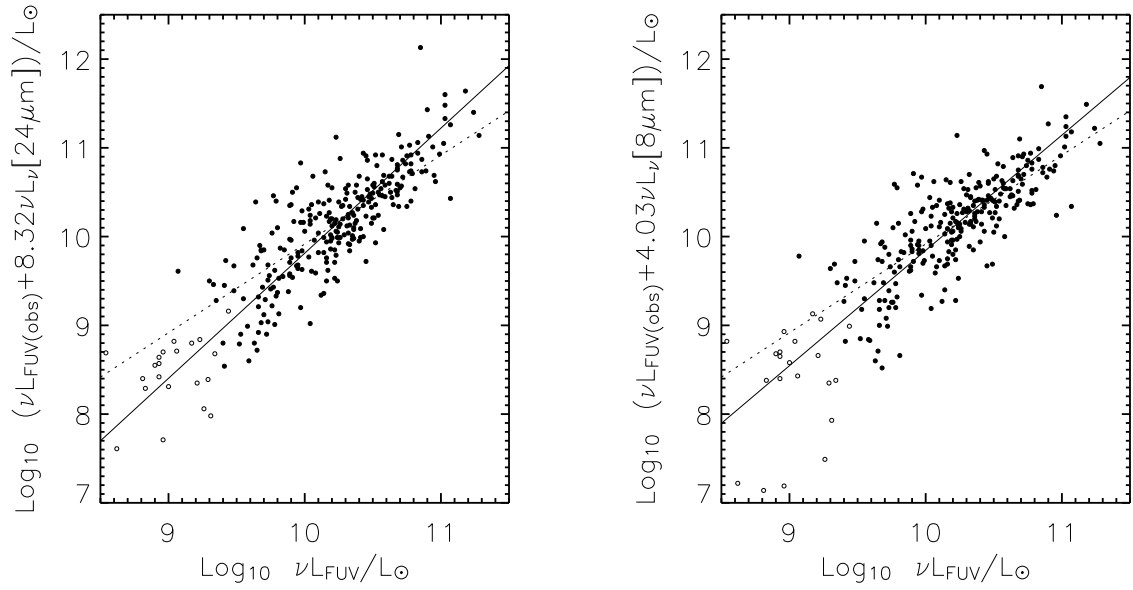}
\caption{Correlations between combined FUV$_{\rm obs}$ $+$ 24$\mu$m \& FUV$_{\rm obs}$ $+$ 8$\mu$m luminosities and
extinction-corrected FUV luminosities for star-forming galaxies. The symbols and line styles are the same as in Figure~\ref{fig8}.}
\label{fig9}
\end{figure}

\begin{figure}
\figurenum{10}
\epsscale{1}
\plotone{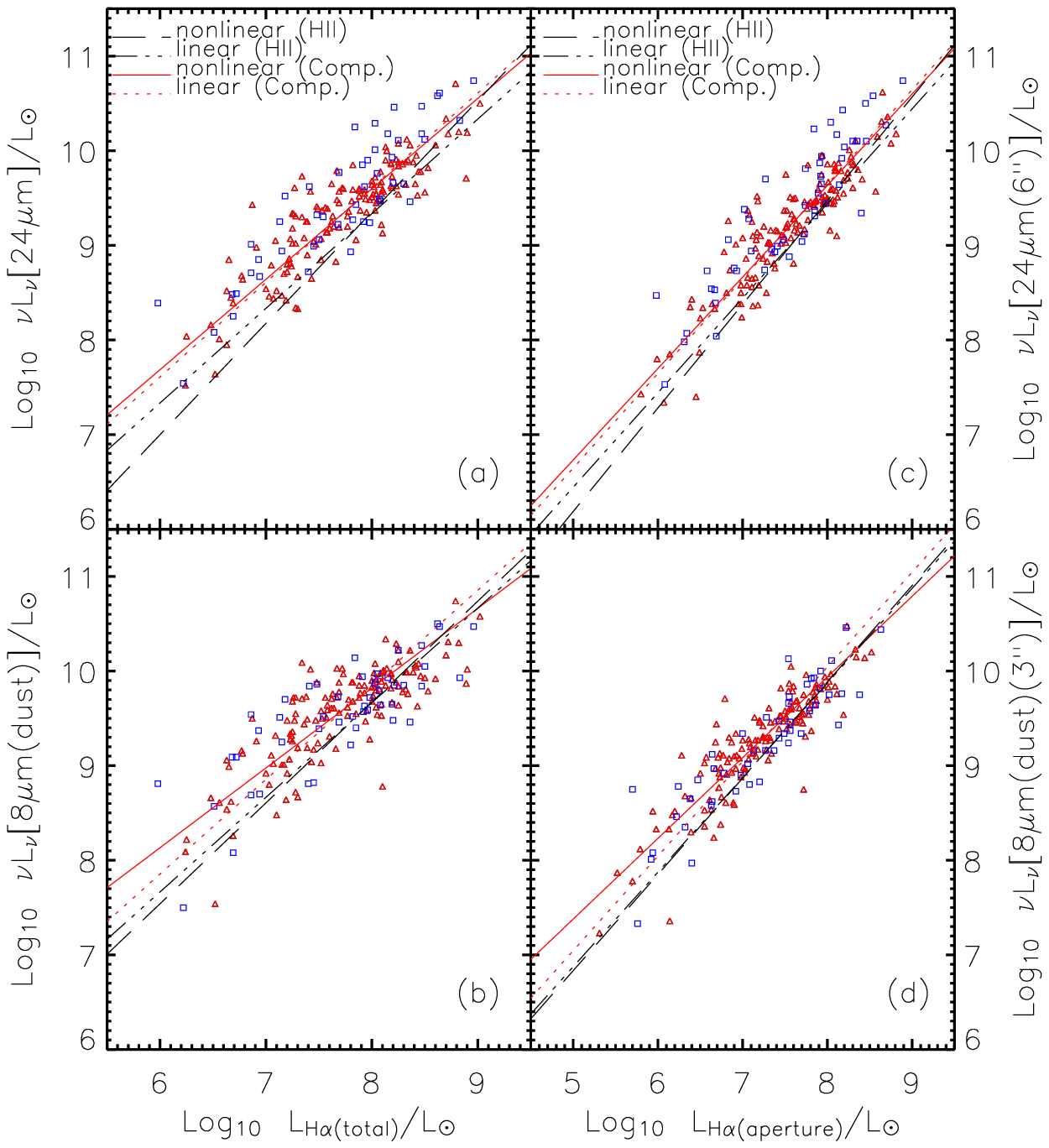}
\caption{Correlations between MIR and H$\alpha$ luminosities for composite galaxies (triangle) and AGNs (box). Panels (a) \& (b): $total$ 
MIR vs. $total$ H$\alpha$ luminosities; Panels (c) \& (d): $aperture$ MIR vs. $aperture$ H$\alpha$ luminosities. The nonlinear and linear fits 
for composite galaxies are illustrated as red solid and dotted lines, while the black dashed and dash-dot lines are the nonlinear and linear 
fits for star-forming galaxies as shown in Figure~\ref{fig4}.}
\label{fig10}
\end{figure}

\begin{figure}
\figurenum{11}
\epsscale{1}
\plotone{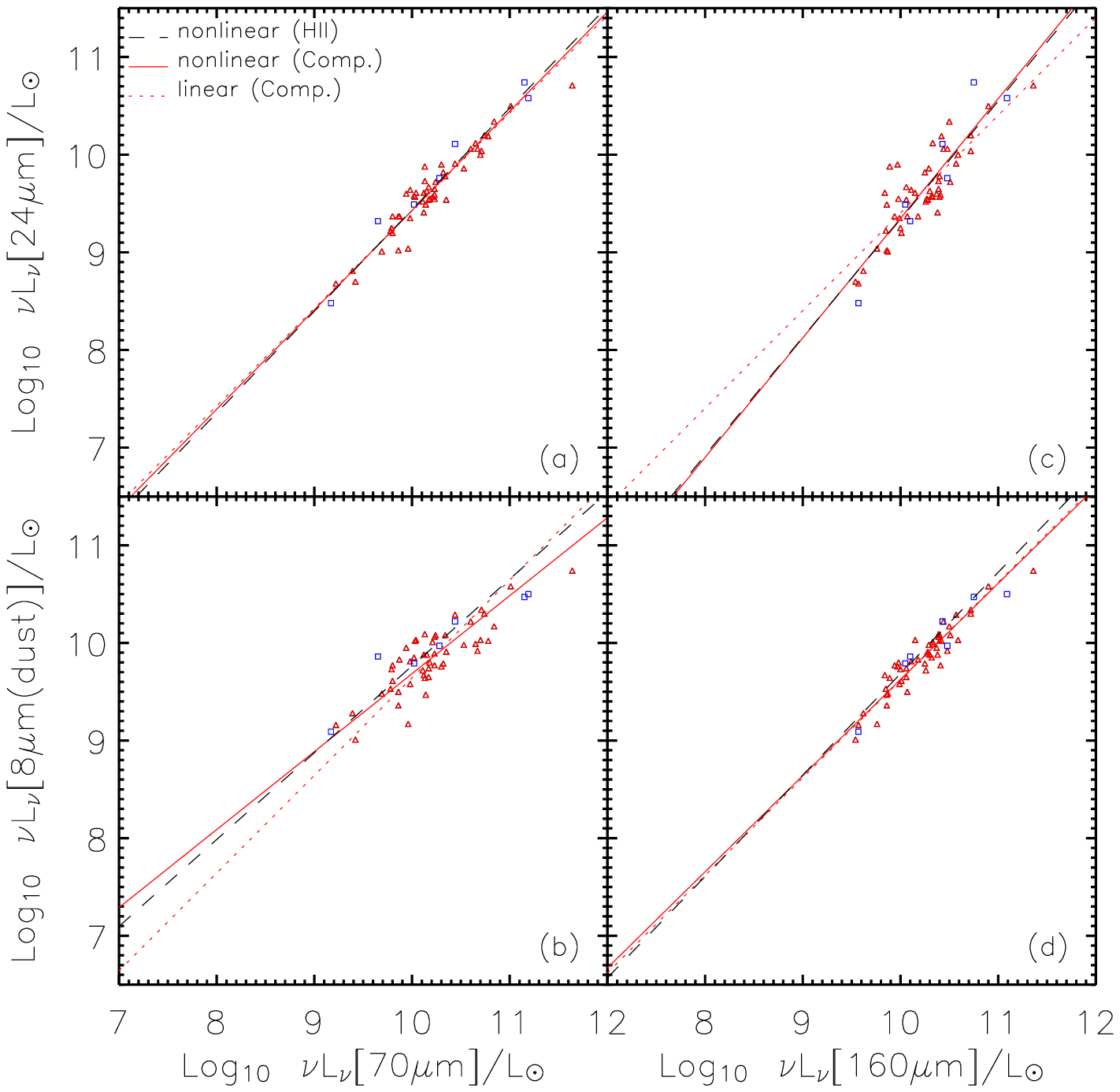}
\caption{Correlations between MIR and FIR luminosities for composite galaxies (triangle) and AGNs (box). The red solid and dotted lines represent 
the nonlinear and linear fits for composite galaxies, while the black dashed lines are the nonlinear fits for star-forming galaxies as shown in 
Figure~\ref{fig6}.}
\label{fig11}
\end{figure}

\begin{figure}
\figurenum{12}
\epsscale{1}
\plotone{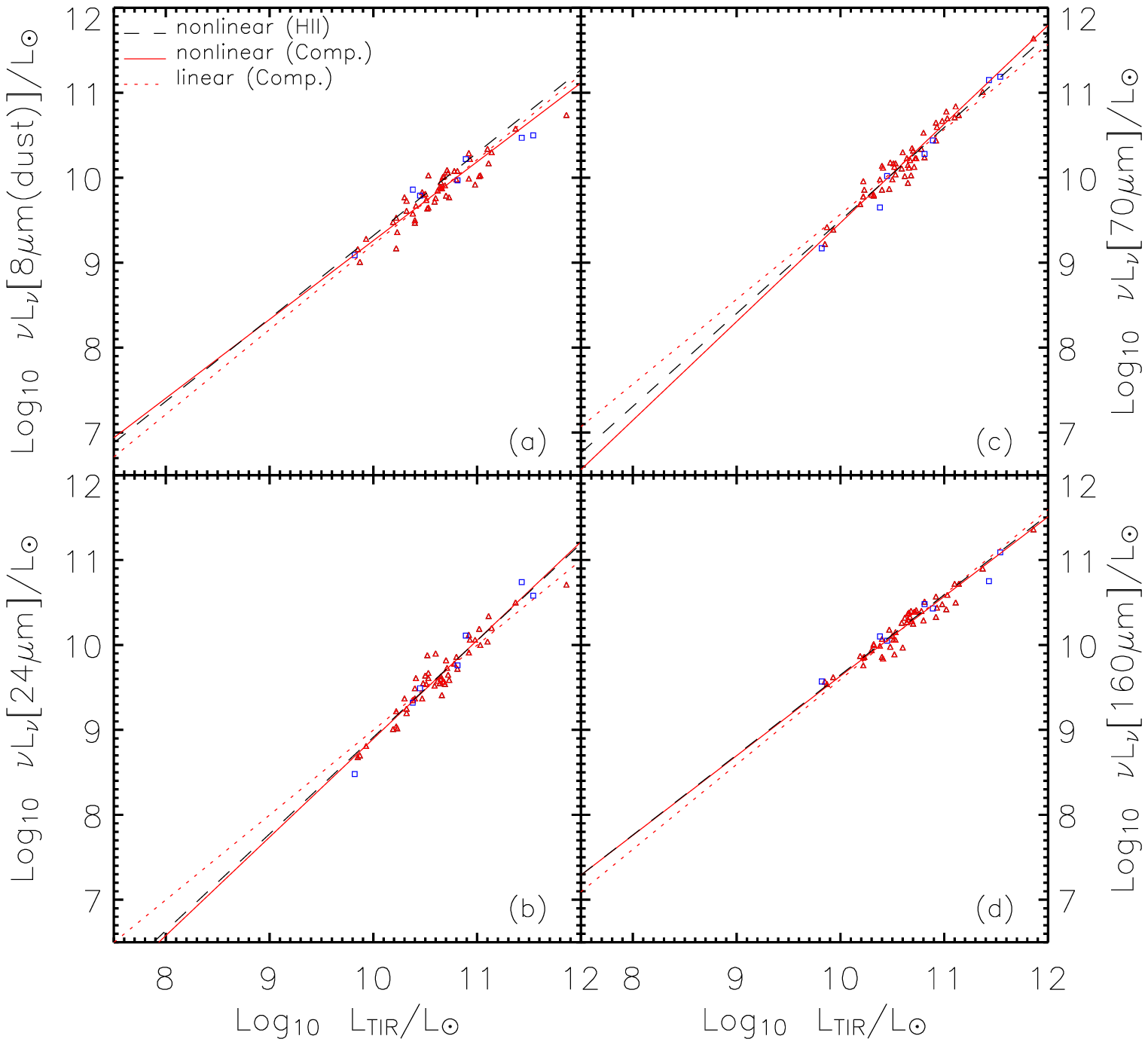}
\caption{Correlations between MIR \& FIR and TIR luminosities for composite galaxies (triangle) and AGNs (box). The red solid and dotted lines 
represent the best nonlinear and linear fits for composite galaxies, while the black dashed lines are the nonlinear fits for star-forming galaxies 
as shown in Figure~\ref{fig7}.}
\label{fig12}
\end{figure}

\begin{figure}
\figurenum{13}
\epsscale{1}
\plotone{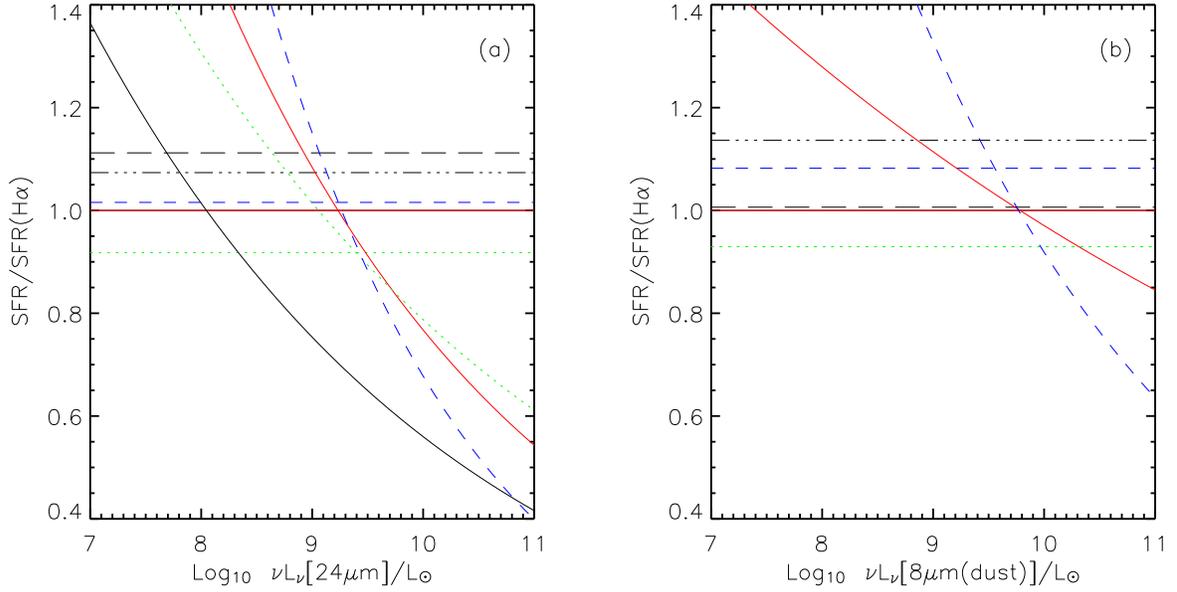}
\caption{SFRs derived from MIR luminosities based on the linear and non-linear correlations between MIR with H$\alpha$ (red solid lines 
for linear, red solid curves for non-linear), FIR (green dotted lines for linear, green dotted curves for non-linear; notice that 
the linear line and non-linear curve are coincident), and UV (blue short-dashed lines for linear, blue short-dashed curves for non-linear) 
luminosities. SFRs derived from MIR luminosities based on the relations from Wu05 between MIR with H$\alpha$ (black long-dashed lines) and 
radio (black dashed-dotted lines) luminosities were also shown in this figure. The black solid curve in the left panel shows the equation 
to compute SFR based on 24$\mu$m luminosity given by \citet{alonso06a} (their equation 3).
}
\label{fig13}
\end{figure}

\begin{figure}
\figurenum{14}
\epsscale{1}
\plotone{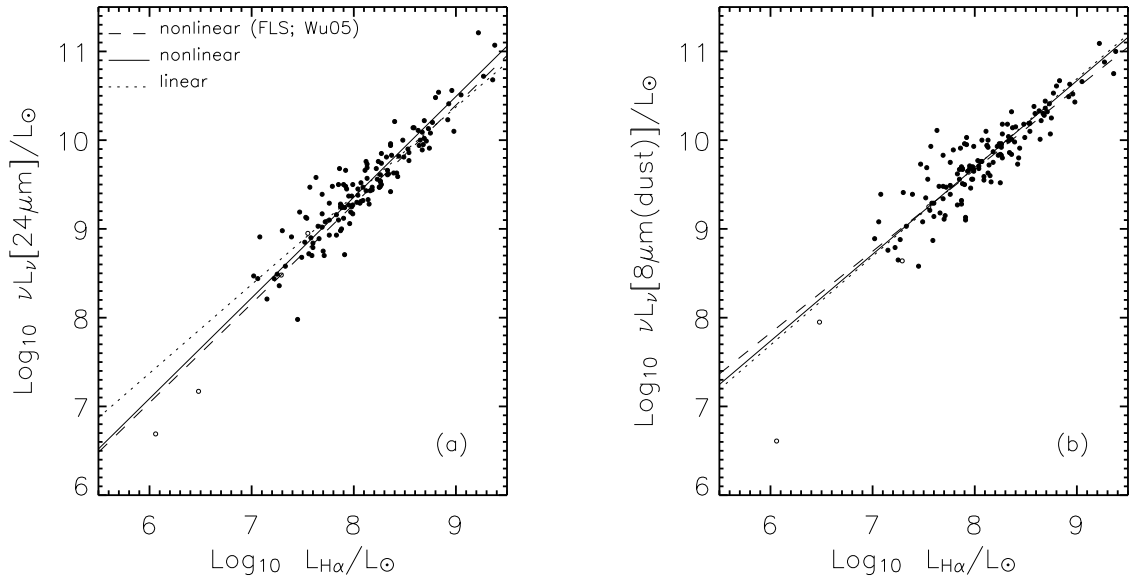}
\caption{Correlations between $total$ MIR and $total$ H$\alpha$ luminosities for galaxies in the FIR sample. The symbols and line styles 
are the same as in Figure~\ref{fig4}.}
\label{fig14}
\end{figure}

Investigating the properties of composite galaxies and AGNs will provide us some information on the origin of MIR emission in 
AGN-hosting galaxies. Figures~\ref{fig8} and~\ref{fig10} show that composite galaxies and AGNs have distinct L(MIR)-L(H$\alpha$) 
and L(MIR)-L(FUV) correlations, i.e., they (especially for the ones with low H$\alpha$ luminosity) have relatively higher 
MIR-to-H$\alpha$/MIR-to-FUV luminosity ratios than star-forming galaxies do. The high MIR-to-H$\alpha$/MIR-to-FUV luminosity 
ratios can not be caused by heavier dust obscuration in the host galaxies of composites and AGNs, since from Figure~\ref{fig3}(c) 
it is clear that the host galaxies of composites and AGNs tend to be earlier Hubble types with redder $u-r$ colors, and it is 
well-known that early-type galaxies should have less gas and dust \citep[e.g.,][]{wen07, li07}. Alternatively, for early-type 
galaxies in which the diffuse IR emissions can not be neglected, their MIR-to-H$\alpha$/MIR-to-FUV luminosity ratios would tend 
to be higher due to additional dust heating by abundant old stellar populations. Furthermore, the higher luminosity (Figure~\ref{fig3}(a)) 
could refer to higher metallicity\citep[e.g.,][]{lamareille04}, while higher metallicity in early-types \citep{wu07, li07} could 
be another explanation to the higher MIR-to-H$\alpha$ luminosity ratios of composite galaxies and AGNs except the re-emission of 
dust heated by AGN.

From Figure~\ref{fig11} it is clear that all the four L(MIR)-L(FIR) correlations for composite galaxies and AGNs are completely 
following those for star-forming galaxies. The L(MIR)-L(TIR) and L(FIR)-L(TIR) correlations for composite galaxies and AGNs are 
also shown to follow those for star-forming galaxies (see Figure~\ref{fig12}). Hence, the contribution from the active nuclear 
to both MIR and FIR emissions is more smaller than the contribution from the HII region, and it could be neglected for those 
weak AGNs. Nevertheless, the similarity of dust heating by AGN or HII region could be another possible explanation. These results 
indicate that we can estimate the total infrared luminosity accurately from monochromatic MIR and FIR luminosities, not only 
for star-forming galaxies (see also $\S$5.3), but also for AGN-hosting galaxies. However, firm conclusion must await a quantitative 
analysis of a large, well-defined, and unbiased sample of FIR-selected galaxies from $AKARI$ and $Herschel$ observations.

\section{Summary}
We present and analyze the correlations between MIR, FIR, TIR, H$\alpha$, and FUV luminosities for a large sample of galaxies 
selected from the $Spitzer$ SWIRE fields. The main results described in this paper can be summarized as follows.

1. The MIR (8$\mu$m[dust] and 24$\mu$m) luminosities of star-forming galaxies are found to be well correlated with their 
H$\alpha$, FIR, TIR, and FUV luminosities. Based on these correlations, we derive corresponding formulae to calculate SFRs 
from L(MIR). All these formulae are well consistent with each other.

2. The linear combination of observed H$\alpha$ (H$\alpha$$_{\rm obs}$) and 24$\mu$m luminosities has tight and linear correlations 
with the extinction-corrected H$\alpha$ luminosities L(H$\alpha$) for star-forming galaxies, so it can be used as a good SFR tracer. 
We determined the scaling parameter $\alpha$ in this combination, and also found the combined H$\alpha$$_{\rm obs}$ $+$ 8$\mu$m(dust) 
luminosities are also well correlated with L(H$\alpha$).

3. Among the four L(MIR)-L(FIR) correlations for star-forming galaxies, the L(24$\mu$m)-L(70$\mu$m) and L(8$\mu$m[dust])-L(160$\mu$m) 
correlations are the tightest and almost linear. The former could be related to young massive star formations, while the latter 
might be relevant to diffuse dust emissions heated by old stellar populations. 

4. Composite galaxies and AGNs have higher MIR-to-H$\alpha$/MIR-to-FUV luminosity ratios than star-forming galaxies, nevertheless 
their correlations among MIR, FIR and TIR luminosities are completely following those for star-forming galaxies.

\acknowledgments
We thank the anonymous referee for constructive comments and suggestions. We acknowledge Drs. X.-Y. Xia, C.-N. Hao, Z.-G. Deng, 
S. Mao, and Y. Shi for advice and helpful discussions, and appreciate J.-S. Huang for kindly providing a serious of SED model 
and Z. Wang, J.-L. Wang, and F.-S. Liu for their capable help and assistance throughout the process of $Spitzer$ data reductions. 
This project is supported by The Ministry of Science and Technology of the People's republic of China through grant 2007CB815406, 
and by NSFC grants 10773014 and 10333060. 

This work is based on observations made with the $Spitzer$ Space Telescope, which is operated by Jet Propulsion Laboratory of the 
California Institute of Technology under NASA Contract 1407. $GALEX$ ($Galaxy Evolution Explorer$) is a NASA Small Explorer, 
launched in April 2003. We gratefully acknowledge NASA's support for construction, operation, and science analysis for the GALEX 
mission, developed in cooperation with the Centre National d'Etudes Spatiales of France and the Korean Ministry of Science and 
Technology. Funding for the SDSS and SDSS-II has been provided by the Alfred P. Sloan Foundation, the Participating Institutions, 
the National Science Foundation, the U.S. Department of Energy, the National Aeronautics and Space Administration, the Japanese 
Monbukagakusho, the Max Planck Society, and the Higher Education Funding Council for England. The SDSS Web Site is http://www.sdss.org/. 
The SDSS is managed by the Astrophysical Research Consortium for the Participating Institutions. The Participating Institutions are the 
American Museum of Natural History, Astrophysical Institute Potsdam, University of Basel, University of Cambridge, Case Western 
Reserve University, University of Chicago, Drexel University, Fermilab, the Institute for Advanced Study, the Japan Participation 
Group, Johns Hopkins University, the Joint Institute for Nuclear Astrophysics, the Kavli Institute for Particle Astrophysics and 
Cosmology, the Korean Scientist Group, the Chinese Academy of Sciences (LAMOST), Los Alamos National Laboratory, the Max-Planck-Institute 
for Astronomy (MPIA), the Max-Planck-Institute for Astrophysics (MPA), New Mexico State University, Ohio State University, University 
of Pittsburgh, University of Portsmouth, Princeton University, the United States Naval Observatory, and the University of Washington.  

\clearpage


\clearpage

\clearpage
 
\clearpage

\clearpage

\clearpage
 
\clearpage

\clearpage

\clearpage

\clearpage

\clearpage

\clearpage

\clearpage
 
\clearpage

\clearpage

\clearpage

%
\begin{deluxetable}{llccc}
\centering
\tablecolumns{5}
\tabletypesize{\footnotesize}
\tablewidth{0pt}
\tablecaption{The expected and observed $NIR-MIR$ colors of blue stars in aperture photometry}
\tablehead{
\colhead{$Field$} & \colhead{$Color$} & \colhead{Expected value\tablenotemark{a}} & \colhead{Observed value} & 
\colhead{Corrected value}}
\startdata
$EN1$  &$K-IRAC1$ & 0.04 & ~~0.09 & ~~0.05 \\
$EN1$  &$K-IRAC2$ & 0.02 &$-$0.03 &$-$0.05 \\
$EN1$  &$K-IRAC3$ & 0.00 &$-$0.03 &$-$0.03 \\
$EN1$  &$K-IRAC4$ & 0.00 & ~~0.07 & ~~0.07 \\
$EN2$  &$K-IRAC1$ & 0.04 & ~~0.06 & ~~0.02 \\
$EN2$  &$K-IRAC2$ & 0.02 &$-$0.03 &$-$0.05 \\
$EN2$  &$K-IRAC3$ & 0.00 &$-$0.03 &$-$0.03 \\
$EN2$  &$K-IRAC4$ & 0.00 & ~~0.07 & ~~0.07 \\
$LH$   &$K-IRAC1$ & 0.04 & ~~0.09 & ~~0.05 \\
$LH$   &$K-IRAC2$ & 0.02 &$-$0.02 &$-$0.04 \\
$LH$   &$K-IRAC3$ & 0.00 &$-$0.02 &$-$0.02 \\
$LH$   &$K-IRAC4$ & 0.00 & ~~0.08 & ~~0.08 \\
\hline
\enddata
\tablenotetext{a}{from \citet{eisenhardt04,lacy05}.}
\label{tab1}
\end{deluxetable}

\clearpage

%
\begin{deluxetable}{lcccc}
\centering
\tablecolumns{5}
\tabletypesize{\footnotesize}
\tablewidth{0pt}
\tablecaption{The number of MIR, FIR and UV sample galaxies with different spectral types}
\tablehead{
\colhead{Sample} & \colhead{$Star-forming$} & \colhead{$Composites$} & 
\colhead{$AGN$} & \colhead{$Total$}}
\startdata
MIR   & 413 & 143 &  49 & 605\\
FIR   & 141 & ~49 &  ~7 & 197\\
UV    & 267 &  65 &  27 & 359\\
\hline
\enddata
\label{tab2}
\end{deluxetable}

\clearpage

%
\begin{deluxetable}{lllrcccrr}
\centering
\rotate
\tablecolumns{9}
\tabletypesize{\footnotesize}
\tablewidth{0pt}
\tablecaption{Correlation coefficients for star-forming galaxies}
\tablehead{
\colhead{Sample} & \colhead{$y$} & \colhead{$x$} & \colhead{$a$} & \colhead{$b$} & 
\colhead{$s$} & \colhead{$r$} & \colhead{$c$} & \colhead{$N$}\\
\colhead{(1)} & \colhead{(2)} & \colhead{(3)} & \colhead{(4)} & \colhead{(5)} & 
\colhead{(6)} & \colhead{(7)} & \colhead{(8)} & \colhead{(9)}}
\startdata
FLS &$\nu L_{\nu}[24\mu m](total)$ &$ L{[H\alpha]}(total)$     &  0.34$\pm$0.19& 1.12$\pm$0.07&  \ldots   & 0.88&  1.29$\pm$0.19& 63\\
FLS &$\nu L_{\nu}[8\mu m(dust)](total)$ &$ L{[H\alpha]}(total)$      &  2.31$\pm$0.15& 0.92$\pm$0.05&  \ldots   & 0.87&  1.68$\pm$0.19& 79\\
MIR &$\nu L_{\nu}[24\mu m](total)$ &$ L{[H\alpha]}(total)$    & $-$0.06$\pm$0.07& 1.18$\pm$0.02& 0.16& 0.92&  1.33$\pm$0.18&379\\
MIR &$\nu L_{\nu}[8\mu m(dust)](total)$ &$ L{[H\alpha]}(total)$     &  1.14$\pm$0.07& 1.07$\pm$0.03& 0.18& 0.88&  1.67$\pm$0.19&379\\
MIR &$\nu L_{\nu}[24\mu m](aper)$ &$ L{[H\alpha]}(aper)$     &  0.63$\pm$0.05& 1.10$\pm$0.02& 0.15& 0.93&  1.37$\pm$0.17&379\\
MIR &$\nu L_{\nu}[8\mu m(dust)](aper)$ &$ L{[H\alpha]}(aper)$      &  1.74$\pm$0.05& 1.02$\pm$0.02& 0.14& 0.94&  1.88$\pm$0.16&379\\
MIR &$L{[H\alpha\_obs/24\mu m]}$  &$L{[H\alpha]}(total)$  &  0.29$\pm$0.04& 0.96$\pm$0.01& 0.09& 0.95&  0.01$\pm$0.11&379\\
MIR &$L{[H\alpha\_obs/8\mu m]}$  &$L{[H\alpha]}(total)$   &  0.86$\pm$0.04& 0.89$\pm$0.01& 0.09& 0.94&  0.01$\pm$0.12&379\\
FIR &$\nu L_{\nu}[24\mu m](total)$ &$ L{[H\alpha]}(total)$    &  0.28$\pm$0.10& 1.13$\pm$0.04& 0.15& 0.92&  1.37$\pm$0.17&137\\
FIR &$\nu L_{\nu}[8\mu m(dust)](total)$ &$ L{[H\alpha]}(total)$     &  1.87$\pm$0.11& 0.98$\pm$0.04& 0.15& 0.87&  1.69$\pm$0.17&137\\
FIR &$\nu L_{\nu}[24\mu m](total)$ &$ \nu L_{\nu}[70\mu m]$  & $-$0.97$\pm$0.05& 1.04$\pm$0.01& 0.07& 0.98& $-$0.56$\pm$0.08&137\\
FIR &$\nu L_{\nu}[8\mu m(dust)](total)$ &$ \nu L_{\nu}[70\mu m]$   &  0.87$\pm$0.07& 0.89$\pm$0.02& 0.10& 0.95& $-$0.24$\pm$0.13&137\\
FIR &$\nu L_{\nu}[24\mu m](total)$ &$ \nu L_{\nu}[160\mu m]$& $-$2.75$\pm$0.10& 1.21$\pm$0.03& 0.12& 0.94& $-$0.63$\pm$0.14&137\\
FIR &$\nu L_{\nu}[8\mu m(dust)](total)$ &$ \nu L_{\nu}[160\mu m]$ & $-$0.69$\pm$0.06& 1.04$\pm$0.02& 0.07& 0.96& $-$0.32$\pm$0.08&137\\
FIR &$\nu L_{\nu}[8\mu m(dust)](total)$ &$ TIR[3-1100\mu m]$&$-$0.44$\pm$0.05&0.98$\pm$0.02& 0.06&0.97&$-$0.70$\pm$0.08&137\\FIR &$\nu L_{\nu}[24\mu m](total)$ &$ TIR[3-1100\mu m]$&$-$2.48$\pm$0.05&1.14$\pm$0.02& 0.07&0.98&$-$1.02$\pm$0.09&137\\
FIR &$\nu L_{\nu}[70\mu m]$ &$ TIR[3-1100\mu m]$&$-$1.45$\pm$0.06&1.09$\pm$0.02& 0.07&0.98&$-$0.45$\pm$0.09&137\\
FIR &$\nu L_{\nu}[160\mu m]$ &$ TIR[3-1100\mu m]$& 0.22$\pm$0.04&0.94$\pm$0.01& 0.05&0.99&$-$0.38$\pm$0.06&137\\
UV  &$\nu L_{\nu}[24\mu m](total)$ &$ FUV$&$-$3.89$\pm$0.14&1.29$\pm$0.04&0.22& 0.86&$-$0.91$\pm$0.24&248\\
UV  &$\nu L_{\nu}[8\mu m(dust)](total)$ &$ FUV$&$-$2.58$\pm$0.14&1.20$\pm$0.04&0.22 & 0.84&$-$0.60$\pm$0.23&248\\
\hline
\enddata
\tablecomments{Col.(1): Sample used for correlation analysis. "FLS" represents the $Spitzer$ First Look Survey sample by Wu05; 
Col.(2)-(3): names of multi-$\lambda$ luminosities. $total$ and $aper$ represent the $total$ and $aperture$ luminosities, and 
all the luminosities are in units of L$_{\odot}$; Col.(4)-(5): the coefficients $a$ and $b$ of the nonlinear fit: 
$\log_{10}(y)=$a$+$b$\log_{10}(x)$; Col.(6): the standard deviation $s$ of the fitting residuals; Col.(7): the coefficient 
$c$ of the linear fit: $\log_{10}(y)=$c$+\log_{10}(x)$; Col.(8): the coefficient $r$ of the Spearman Rank-order correlation 
analysis; Col.(9): the number of sample galaxies (after excluding dwarf galaxies) used for the fitting procedures.}
\label{tab3}
\end{deluxetable}

\clearpage

%
\begin{deluxetable}{lllrcccrr}
\centering
\rotate
\tablecolumns{9}
\tabletypesize{\footnotesize}
\tablewidth{0pt}
\tablecaption{Correlation coefficients for composite galaxies}
\tablehead{
\colhead{Sample} & \colhead{$y$} & \colhead{$x$} & \colhead{$a$} & \colhead{$b$} & 
\colhead{$s$} & \colhead{$r$} & \colhead{$c$} & \colhead{$N$}\\
\colhead{(1)} & \colhead{(2)} & \colhead{(3)} & \colhead{(4)} & \colhead{(5)} & 
\colhead{(6)} & \colhead{(7)} & \colhead{(8)} & \colhead{(9)}}
\startdata
MIR &$\nu L_{\nu}[24\mu m](total)$ &$ L{[H\alpha]}(total)$    & 1.95$\pm$0.11& 0.96$\pm$0.04& 0.19& 0.87&  1.61$\pm$0.22&143\\
MIR &$\nu L_{\nu}[8\mu m](total)$ &$ L{[H\alpha]}(total)$     & 3.07$\pm$0.12& 0.84$\pm$0.04& 0.21& 0.79&  1.86$\pm$0.27&143\\
MIR &$\nu L_{\nu}[24\mu m](aper)$ &$ L{[H\alpha]}(aper)$    & 1.90$\pm$0.10& 0.97$\pm$0.04& 0.18& 0.89&  1.64$\pm$0.20&143\\
MIR &$\nu L_{\nu}[8\mu m](aper)$ &$ L{[H\alpha]}(aper)$     & 3.11$\pm$0.08& 0.85$\pm$0.03& 0.15& 0.89&  2.04$\pm$0.21&143\\
FIR &$\nu L_{\nu}[24\mu m]$ &$ \nu L_{\nu}[70\mu m]$  &$-$0.75$\pm$0.15& 1.02$\pm$0.05& 0.09& 0.90& $-$0.57$\pm$0.10&49\\
FIR &$\nu L_{\nu}[8\mu m]$ &$ \nu L_{\nu}[70\mu m]$   & 1.70$\pm$0.21& 0.80$\pm$0.07& 0.14& 0.78& $-$0.36$\pm$0.19&49\\
FIR &$\nu L_{\nu}[24\mu m]$ &$ \nu L_{\nu}[160\mu m]$&-2.90$\pm$0.31& 1.22$\pm$0.09& 0.16& 0.78& $-$0.60$\pm$0.19&49\\
FIR &$\nu L_{\nu}[8\mu m]$ &$ \nu L_{\nu}[160\mu m]$ &-0.22$\pm$0.16& 0.98$\pm$0.05& 0.09& 0.92& $-$0.38$\pm$0.09&49\\
FIR &$\nu L_{\nu}[8\mu m]$ &$ TIR[3-1100\mu m]$&$-$0.02$\pm$0.17&0.93$\pm$0.05& 0.10&0.91&$-$0.79$\pm$0.11&49\\
FIR &$\nu L_{\nu}[24\mu m]$ &$ TIR[3-1100\mu m]$&$-$2.69$\pm$0.17&1.16$\pm$0.05& 0.09&0.90&$-$1.00$\pm$0.11&49\\
FIR &$\nu L_{\nu}[70\mu m]$ &$ TIR[3-1100\mu m]$&$-$2.15$\pm$0.15&1.16$\pm$0.04& 0.08&0.93&$-$0.43$\pm$0.10&49\\
FIR &$\nu L_{\nu}[160\mu m]$ &$ TIR[3-1100\mu m]$&0.26$\pm$0.12&0.94$\pm$0.04& 0.07&0.95&$-$0.41$\pm$0.09&49\\
\hline
\enddata
\tablecomments{The definition of columns are the same as in Table~\ref{tab3}.}
\label{tab4}
\end{deluxetable}


\begin{thebibliography}{}
\bibitem[Alonso-Herrero et al.(2006a)]{alonso06a} Alonso-Herrero, A., Rieke, G. H., Rieke, M. J., Colina, P{\'e}rez-Gonz{\'a}lez, P. G., \& Ryder, S. D.\ 2006, \apj, 650, 835
\bibitem[Alonso-Herrero et al.(2006b)]{alonso06b} Alonso-Herrero, A., Colina, L., Packham, C., Díaz-Santos, T., Rieke, G. H., Radomski, J. T., \& Telesco, C. M.\ 2006, \apj, 652, 83
\bibitem[Baldwin et al.(1981)]{baldwin81} Baldwin, J. A., Phillips, M. M., \& Terlevich, R.\ 1981, \pasp, 93, 5B
\bibitem[Beichman(1987)]{beichman87} Beichman, C. A.\ 1987, \araa, 25, 521
\bibitem[Bell(2003)]{bell03} Bell, E. F.\ 2003, \apj, 586, 794
\bibitem[Bell et al.(2005)]{bell05} Bell, E.~F., et al.\ 2005, \apj, 625, 23 
\bibitem[Bertin \& Arnouts(1996)]{bertin96} Bertin, E. \& Arnouts, S. 1996. \aaps, 117, 393
\bibitem[Bessel \& Brett(1988)]{bessel88} Bessel. M. S. \& Brett, J. M.\ 1988, \pasp, 100, 1134
\bibitem[Blanton et al.(2003)]{blanton03} Blanton, M. R., et al.\ 2003, \aj, 125, 2348
\bibitem[Blanton \& Roweis(2007)]{blanton07} Blanton, M. R. \& Roweis, S.\ 2007, \aj, 133, 734
\bibitem[Boselli et al.(2004)]{boselli04} Boselli, A., Lequeux, J., \& Gavazzi, G.\ 2004, \aap, 428, 409
\bibitem[Buat et al.(2005)]{buat05} Buat, V., et al.\ 2005, \apj, 619, L51
\bibitem[Burgarella et al.(2005)]{burgarella05} Burgarella, D., Buat, V., \& Iglesias-P\'aramo. J.\ 2005, \mnras, 360, 1413
\bibitem[Cardelli et al.(1989)]{cardelli89} Cardelli, J. A., Clayton, G. C., \& Mathis, J. S.\ 1989, \apj, 345, 245
\bibitem[Calzetti(2001)]{calzetti01} Calzetti, D.\ 2001, \pasp, 113, 1449
\bibitem[Calzetti et al.(2005)]{calzetti05} Calzetti, D., et al.\ 2005, \apj, 633, 871
\bibitem[Calzetti et al.(2007)]{calzetti07} Calzetti, D., et al.\ 2007, \apj, 666, 870
\bibitem[Cao \& Wu (2007)]{cao07} Cao, C., \& Wu, H.\ 2007, \aj, 133, 1710
\bibitem[Cao et al.(2008)]{cao08} Cao, C., Wu, H., Wang, Z., Ho, L.~C., Huang, J.~S., \& Deng, Z.~G.\ 2008, New Astronomy, 13, 16 
\bibitem[Choi(2006)]{choi06} Choi, P. I. 2006, \apj, 637, 227
\bibitem[Cortese et al.(2006)]{cortese06} Cortese, L., et al. 2006, \apj, 637, 242
\bibitem[Cutri et al.(2003)]{cutri03} Cutri, R. M., et al.\ 2003, The IRSA 2MASS All-Sky Point Source Catalog, NASA/IPAC
\bibitem[Daddi et al.(2007)]{daddi07} Daddi, E., et al.\ 2007, \apj, 670, 173 
\bibitem[Dale \& Helou (2002)]{dale02} Dale, D. A., \& Helou, G.\ 2002, \apj, 576, 159
\bibitem[Draine(2003)]{draine03} Draine, B.~T.\ 2003, \araa, 41, 241 
\bibitem[Elbaz et al.(2002)]{elbaz02} Elbaz, D., Cesarsky, C. J., Chanial, P., Aussel, H., Franceschini, A., Fadda, D., \& Chary, R. R.\ 2002, \aap, 384, 848
\bibitem[Elbaz et al.(2007)]{elbaz07} Elbaz, D., et al.\ 2007, \aap, 468, 33
\bibitem[Eisenhardt et al.(2004)]{eisenhardt04} Eisenhardt, P. R., et al.\ 2004, \apjs, 154, 48
\bibitem[Engelbracht et al.(2005)]{engelbracht05} Engelbracht, C. W., Gordon, K. D., Rieke, G. H., Werner, M. W., Dale, D. A., \& Latter, W. B.\ 2005, \apj, 628, L29
\bibitem[Engelbracht et al.(2008)]{engelbracht08} Engelbracht, C. W., Rieke, G. H., Gordon, K. D., Smith, J.-D. T., Werner, M. W., Moustakas, J., Willmer, C. N. A., \& Vanzi, L.\ 2008, in press (astro-ph/0801.1700)
\bibitem[Fazio et al.(2004)]{fazio04} Fazio, G. G., et al.\ 2004, \apjs, 154, 10
\bibitem[Flores et al.(2004)]{flores04} Flores, H., Hammer, F., Elbaz, D., Cesarsky, C. J., Liang, Y. C., Fadda, D., \& Gruel, N.\ 2004, \aap, 415, 885
\bibitem[F\"orster Schreiber et al.(2004)]{schreiber04} F\"orster Schreiber, N. M., Roussel, H., Sauvage, M., \& Charmandaris, V.\ 2004, \aap, 419, 501
\bibitem[Galliano et al.(2005)]{galliano05} Galliano, F., Madden, S. C., Jones, A. P., Wilson, C. D., \& Bernard, J.-P.\ 2005, \aap, 434, 867G
\bibitem[Gil de Paz et al.(2007)]{gil07} Gil de Paz, A., et al.\ 2007, \apjs, 173, 185 
\bibitem[Gillett et al.(1973)]{gillett73} Gillett, F. C., Forrest, W. J., \& Merrill, K.\ 1973, \apj, 183,87
\bibitem[Gordon et al.(2004)]{gordon04} Gordon, K.~D., et al.\ 2004, \apjs, 154, 215 
\bibitem[Gu et al.(2006)]{gu06} Gu, Q.-S., Zhao, Y.-H., Shi, L., Peng, Z.-X., \& Luo, X.-L.\ 2006, \aj, 131, 806
\bibitem[Heckman(1980)]{heckman80} Heckman, T. M.\ 1980, \aap, 87, 152
\bibitem[Helou et al.(2004)]{helou04} Helou, G., et al.\ 2004, \apjs, 154, 253
\bibitem[Ho et al.(1997)]{ho97} Ho, L. C., Fillipenko, A. V., \& Sargent, W. L. W.\ 1997, \apj, 487, 568
\bibitem[Hollenbach \& Tielens (1997)]{hollenbach97} Hollenbach, D. J., \& Tielens, A. G. G. M.\ 1997, \araa, 35, 179
\bibitem[Hopkins et al.(2003)]{hopkins03} Hopkins, A. M., et al.\ 2003, \apj, 599, 971
\bibitem[Huang et al.(2004)]{huang04} Huang, J.-S., et al.\ 2004, \apjs, 154, 44
\bibitem[Huang et al.(2007)]{huang07} Huang, J.-S., et al.\ 2007, \apj, 664, 840 
\bibitem[Hunter et al.(1986)]{hunter86} Hunter, D. A., Gillett, F. C., Gallagher, J. S., Rice, W. L., \& Low, F. J.\ 1986, \apj, 303, 171
\bibitem[Jonsson(2004)]{jonsson04} Jonsson, P.\ 2004, Ph.D. thesis, Univ. California, Santa Cruz
\bibitem[Kauffmann et al.(2003)]{kauffmann03} Kauffmann, G., et al.\ 2003, \mnras, 346, 1055
\bibitem[Kennicutt et al.(1994)]{kennicutt94} Kennicutt, R.~C., Jr., Tamblyn, P., \& Congdon, C. E.\ 1994, \apj, 435, 22
\bibitem[Kennicutt(1998)]{kennicutt98} Kennicutt, R.~C., Jr.\ 1998, \araa, 36, 189
\bibitem[Kennicutt et al.(2003)]{kennicutt03} Kennicutt, R.~C., Jr., et al.\ 2003, \pasp, 115, 928
\bibitem[Kennicutt et al.(2007)]{kennicutt07} Kennicutt, R.~C., Jr., et al.\ 2007, \apj, 671, 333
\bibitem[Kessler et al.(1996)]{kessler96} Kessler, M. F., et al.\ 1996, \aap, 315, L27 
\bibitem[Kewley et al.(2001)]{kewley01} Kewley, L. J., Heisler, C. A., Dopita, M. A., \&  Lumsden, S.\ 2001, \apjs, 132, 37
\bibitem[Kewley et al.(2006)]{kewley06} Kewley, L. J., Groves, B., Kauffmann, G., \& Heckman, T. M.\ 2006, \mnras,372,961 
\bibitem[Kong et al.(2004)]{kong04} Kong, X., Charlot, S., Brinchmann, J., \& Fall, S. M.\ 2004, \mnras, 349, 769
\bibitem[Lacy et al.(2005)]{lacy05} Lacy, M., et al.\ 2005, \apjs, 161, 41
\bibitem[Lamareille et al.(2004)]{lamareille04}Lamareille, F., Mouhcine, M., Contini, T., Lewis, I., \& Maddox, S.\ 2004, \mnras, 350, 396
\bibitem[Lehnert \& Heckman (1996)]{lehnert96} Lehnert, M. D, \& Heckman, T. M.\ 1996, \apj, 472, 546
\bibitem[L\'eger \& Puget (1984)]{leger84} L\'eger, A., \& Puget, J. L.\ 1984, \aap, 137, L5
\bibitem[Leitherer \& Heckman (1995)]{leitherer95} Leitherer, C., \& Heckman, T. M.\ 1995, \apjs, 96, 9
\bibitem[Leitherer et al.(1999)]{leitherer99} Leitherer, C., et al.\ 1999, \apjs, 123, 3
\bibitem[Li \& Draine (2002)]{li02} Li, A., \& Draine, B.~T.\ 2002, \apj, 572, 232 
\bibitem[Li et al.(2007)]{li07} Li, H.-N., Wu, H., Cao, C., \& Zhu, Y.-N.\ 2007, \aj, 134, 1315 
\bibitem[Lisker et al.(2006)]{lisker06} Lisker, T., Glatt, K., Westera, P., \& Grebel, E. K.\ 2006, \aj, 132, 2432
\bibitem[Lonsdale et al.(2003)]{lonsdale03} Lonsdale, C. J., et al.\ 2003, \pasp, 115, 897
\bibitem[Madden(2000)]{madden00} Madden, S.~C.\ 2000, New Astronomy Review, 44, 249 
\bibitem[Marcillac et al.(2006)]{marcillac06} Marcillac, D., Elbaz, D., Charlot, S., Liang, Y. C., Hammer, F., Flores, H., Cesarsky, C., \& Pasquali, A.\ 2006, \aap, 458, 369M
\bibitem[Martin et al.(2005)]{martin05} Martin, D. C., et al.\ 2005, \apj, 619, L1
\bibitem[Peeters et al.(2004)]{peeters04} Peeters, E., Spoon, H. W. W., \& Tielens, A. G. G. M.\ 2004, \apj, 613, 986
\bibitem[P{\'e}rez-Gonz{\'a}lez et al.(2006)]{perez06} P{\'e}rez-Gonz{\'a}lez, P. G., et al.\ 2006, \apj, 648, 987
\bibitem[Popescu et al.(2005)]{popescu05} Popescu, C.~C., et al.\ 2005, \apjl, 619, L75 
\bibitem[Prescott et al.(2007)]{prescott07} Prescott, M.~K.~M., et al.\ 2007, \apj, 668, 182 
\bibitem[Puget \& L\'eger(1989)]{puget89} Puget, J. L., \& L\'eger, A.\ 1989, \araa, 27, 161
\bibitem[Rieke et al.(2004)]{rieke04} Rieke, G.~H., et al.\ 2004, \apjs, 154, 25
\bibitem[Roussel et al.(2001)]{roussel01} Roussel, H., Sauvage, M., Vigroux, L., \& Bosma, A.\ 2001, \aap, 372, 427
\bibitem[Salim et al.(2005)]{salim05}Salim,S., et al.\ 2005, \apj, 619, 39 
\bibitem[Salim et al.(2007)]{salim07}Salim,S., et al.\ 2007, \apjs, 173, 267
\bibitem[Sanders \& Mirabel (1996)]{sanders96} Sanders, D. B.,\& Mirabel, I. F.\ 1996, \araa, 34, 749
\bibitem[Seibert et al.(2005)]{seibert05} Seibert, M., et al.\ 2005, \apj, 619, L55
\bibitem[Siebenmorgen et al.(2004)]{siebenmorgen04} Siebenmorgen, R., Kr{\"u}gel, E., \& Spoon, H.~W.~W.\ 2004, \aap, 414, 123
\bibitem[Siebenmorgen \& Kr{\"u}gel(2007)]{siebenmorgen07} Siebenmorgen, R., \& Kr{\"u}gel, E.\ 2007, \aap, 461, 445 
\bibitem[Smith et al.(2002)]{smith02} Smith, J. A., et al.\ 2002, \aj, 123, 2121
\bibitem[Smith et al.(2007)]{smith07} Smith, J.~D.~T., et al.\ 2007, \apj, 656, 770 
\bibitem[Strauss et al.(2002)]{strauss02} Strauss, M. A., et al.\ 2002, \aj, 124, 1810
\bibitem[Stoughton et al.(2002)]{stoughton02} Stoughton, C., et al.\ 2002, \aj, 123, 485
\bibitem[Takeuchi et al.(2005)]{takeuchi05} Takeuchi, T. T., Buat, V., Iglesias-Páramo, J., Boselli, A., \& Burgarella, D.\ 2005, \aap, 432,423
\bibitem[Thilker et al.(2007)]{thilker07} Thilker, D. A., et al. \ 2007, \apjs, 173, 538 
\bibitem[Thuan \& Martin (1981)]{thuan81} Thuan, T. X., \& Martin, G. E.\ 1981, \apj, 247, 823
\bibitem[Tremonti et al.(2004)]{tremonti04} Tremonti, C. A., et al.\ 2004, \apj, 613, 898
\bibitem[Treyer et al.(2007)]{treyer07} Treyer, M., et al.\ 2007, \apjs, 173, 256
\bibitem[Veilleux \& Osterbrock (1987)]{veilleux87} Veilleux, S., \& Osterbrock, D. E.\ 1987, \apjs, 63, 295
\bibitem[Verma et al.(2005)]{verma05} Verma, A., Charmandaris, V., Klaas, U., Lutz, D., \& Haas, M.\ 2005, Space Science Reviews, 119, 355
\bibitem[Weedman et al.(2005)]{weedman05} Weedman, D. W., et al.\ 2005, \apj, 633, 706
\bibitem[Wen et al.(2007)]{wen07} Wen, X.-Q., Wu, H., Cao, C., Xia, X.-Y.\ 2007, Chinese Journal of Astronomy and Astrophysics, 7, 187
\bibitem[Werner et al.(2004)]{werner04} Werner, M. W., et al.\ 2004, \apjs, 154, 1
\bibitem[Willner et al.(1977)]{willner77} Willner, S. P., Soifer, B. T., Russell, R. W., Joyce, R. R., \& Gillett, F. C.\ 1977, \apj, 217, 121
\bibitem[Wu et al.(1998)]{wu98} Wu, H., Zou, Z. L., Xia, X. Y., \& Deng, Z. G.\ 1998, \aaps, 132, 181
\bibitem[Wu et al.(2005)]{wu05} Wu, H., Cao, C., Hao, C.-N., Liu, F.-S., Wang, J.-L., Xia, X.-Y., Deng, Z.-G., \& Young, C.~K. S.\ 2005, \apjl, 632, L79
\bibitem[Wu et al.(2007)]{wu07} Wu, H., Zhu, Y. N., Cao, C., \& Qin, B.\ 2007, \apj, 668, 87
\bibitem[Wu et al.(2006)]{wyl06} Wu, Y., Charmandaris, V., Hao, L., Brandl, B.~R., Bernard-Salas, J., Spoon, H.~W.~W., \& Houck, J.~R.\ 2006, \apj, 639, 157 
\bibitem[York et al.(2000)]{york00} York, D. G., et al.\ 2000, \aj, 120, 1579 
\bibitem[Zamojski et al.(2007)]{zamojski07} Zamojski, M. A., et al.\ 2007, \apjs, 172, 468

\end{thebibliography}
\end{document}